\documentclass[bibyear]{aa}
\usepackage{amsmath}
\usepackage{graphicx}
\usepackage{txfonts}
\usepackage{amssymb}
\usepackage[]{natbib}
\usepackage{mathrsfs}
\usepackage{caption}
\usepackage{float}
\usepackage{afterpage}
\usepackage{amstext}
\usepackage{subfigure}
\newcommand{\bjdtdb}[0]{\mbox{$\mathrm{BJD}_{\mathrm{TDB}}$}}

\begin{document}
   \title{Modelling systematics of ground-based transit photometry I.}

   \subtitle{Implications on transit timing variations}

   \author{C. von Essen$^{1,5}$, S. Cellone$^{2,3,4}$, M. Mallonn$^{6}$, B. Tingley$^{1}$, M. Marcussen$^{7}$}
   \institute{$^1$Stellar Astrophysics Centre, Department of Physics and Astronomy, Aarhus University, Ny Munkegade 120, DK-8000 Aarhus C, Denmark\\
           $^2$Consejo Nacional de Investigaciones Cient\'{\i}ficas y T\'ecnicas, Godoy Cruz 2290, C1425FQB, Ciudad Aut\'onoma de Buenos Aires, Argentina\\
           $^3$Facultad de Ciencias Astron\'omicas y Geof\'{\i}sicas, Universidad Nacional de La Plata, Paseo del Bosque, B1900FWA, La Plata, Argentina\\
           $^4$Instituto de Astrof\'{\i}sica de La Plata (CCT-La Plata, CONICET-UNLP), Paseo del Bosque, B1900FWA, La Plata, Argentina\\
           $^5$Hamburger Sternwarte, Universit\"at Hamburg, Gojenbergsweg 112, 21029 Hamburg, Germany\\
           $^6$Leibniz-Institut f\"ur Astrophysik Potsdam, An der Sternwarte 16, 14482, Potsdam, Germany\\
           $^7$Department of Physics and Astronomy, Aarhus University, Ny Munkegade 120, DK-8000 Aarhus C, Denmark}
   \authorrunning{C. von Essen et al.}
   \date{July 2016}

\abstract{} 
{The transit timing variation technique (TTV) has been widely used to
  detect and characterize multiple planetary systems. Due to the
  observational biases imposed mainly by the photometric conditions
  and instrumentation and the high signal-to-noise required to produce
  primary transit observations, ground-based data acquired using small
  telescopes limit the technique to the follow-up of hot
  Jupiters. However, space-based missions such as {{\it Kepler}} and
  {{\it CoRoT}} have already revealed that hot Jupiters are mainly
  found in single systems. Thus, it is natural to question ourselves
  if we are properly using the observing time at hand carrying out
  such follow-ups, or if the use of medium-to-low quality transit
  light curves, combined with current standard techniques of data
  analysis, could be playing a main role against exoplanetary search
  via TTVs. The purpose of this work is to investigate to what extent
  ground-based observations treated with current modelling techniques
  are reliable to detect and characterize additional planets in
  already known planetary systems.}
{To meet this goal, we simulated typical primary transit observations
  of a hot Jupiter mimicing an existing system, Qatar-1. To resemble
  ground-based observations we attempt to reproduce, by means of
  physically and empirically motivated relationships, the effects
  caused by the Earth's atmosphere and the instrumental setup on
  the synthetic light curves. Therefore, the synthetic data present
  different photometric quality and transit coverage. In addition, we
  introduced a perturbation in the mid-transit times of the hot
  Jupiter, caused by an Earth-sized planet in a 3:2 mean motion
  resonance. Analyzing the synthetic light curves produced after
  certain epochs, we attempt to recover the synthetically added TTV
  signal by means of usual primary transit fitting techniques.}
{In this work we present an extensive description of the noise sources
  accounted for that are usually associated to ground-based
  observations, along with a discussion and motivation for their
  consideration. Additionally, we provide a comparison analysis
  between real and synthetic light curves, to test up to what extent
  do both data sets present the same degree of distortion. Finally, we
  show how standard techniques recover (or not) the TTV signal, and
  determine a light curve ``quality factor'' that would be needed to
  properly recover the TTVs.}
 {}

   \maketitle  


\section{Introduction}

The advent of highly-accurate long-term and space-based observations
such as Kepler \citep{Borucki2010} and CoRoT \citep{Baglin2006} marked
a new era for exoplanet search. For instance, Kepler light curves
already revealed clear signatures of transit timing variations
\citep[TTVs; see
  e.g.,][]{Holman2010,Lissauer2011,Ballard2011,Steffen2013}, a
technique that relies on the variations in the timings of transits to
detect and characterize planetary systems with members that can be as
light as one Earth mass or below \citep{Agol2005,Holman2005}. Despite
their indisputable power, these space missions were neither designed
to observe the whole sky nor to follow up already known single
exoplanetary systems outside their fields of view. At present, this
role can only be played by ground-based telescopes located across the
globe.

To produce reliable TTV studies, optimal ground-based observations of
primary transits would require, to begin with, a sufficiently long
time baseline, good phase coverage, and deep primary
transits. However, TTV studies are carried out under less strict
conditions. Literature already reveals how misleading can be
ground-based observations when orbital and physical parameters derived
from them are being compared to each other. For instance, after
analyzing archival data plus two extra consecutive transit
observations, \cite{Diaz2008} reported TTVs in OGLE-Tr-111. Later on,
6 additional transit light curves and a new re-analysis of the
complete data set revealed no detection of TTVs
\citep{Adams2010}. Another system that has been systematically
observed during primary transit is WASP-3. Using observations obtained
by means of small aperture telescopes, \cite{Maciejewski2010} firstly
reported the detection of TTVs in the system. Additionally, after
collecting more than 3 years of transit observations \cite{Eibe2012}
reported probable variations in the transit duration, instead of the
claimed TTVs. However, \cite{Monatalto2012} studied thirty-eight
archival light curves in an homogeneous way, and found no significant
evidence of TTVs in the system. Also our group encountered
  problems in identifying TTVs: although \cite{vonEssen2013} reported
  indications of TTVs in the Qatar-1 system, \cite{Maciejewski2015}
  and \cite{Mislis2015} did not reproduce them using more precise and
  extensive data. Even variations in the inclination were wrongly
  claimed. Before the Kepler team released the first quarters,
\cite{Mislis2010} reported a significant variation in the inclination
of \mbox{TrES-2}, one planetary system within Kepler's field of
view. Afterwards, \cite{Schroeter2012} re-analyzed all the published
observations in addition to the Kepler data, finding that while
ground-based observations revealed a declining trend in inclination,
Kepler data were consistent with no variation at all
\citep[see][Figure 2]{Schroeter2012}. Intriguingly, TrES-2b produces
one of the largest primary transit depths and the host star is
relatively bright, which would make it an easy target to be observed
from the ground.

Although the TTV technique is a powerful method to detect exoplanets
in multiple systems, the systematic disagreement between authors
causes critical readers to disbelieve low-amplitude results. Added to
this, planet formation theories \citep{Fogg2007,Mandell2007} and
highly precise observations \citep{Steffen2012,Steffen2013b} reveal
that hot Jupiters are prone to conform single systems instead of
multiple ones. It is natural then to ask ourselves if standard
techniques used to analyze ground based transit data are reliable
enough to produce robust results, or if the technology used to carry
out these observations plus the effects introduced by our atmosphere
on photometric data are playing a main role against us. These
circumstances motivated us to write a code capable to create realistic
synthetic light curves affected by systematics commonly present in
ground-based observations. The main goal is to study under which
conditions can the artificially added TTV signal be retrieved. In this
work we present a detailed description of our code, of the noise
sources that are injected into the light curves, and we show a
rigorous test that quantifies the resemblance between our synthetic
light curves and real data. We show how and to which extent can
systematics not properly accounted for reproduce TTV signals and
quantify their impact over the characterization of the perturbing
planet. We finish our work trying to characterize light curve
observables that would be associated to reliable mid-transit times.

\section{Our code: Generalities}

\subsection{Starting point: Stellar and planetary properties}

To begin with, our code needs the configuration and the properties of
the system to be simulated. In the case of the host star, the inputs
are the stellar radius, $R_{\mathrm{S}}$, the spectral and
sub-spectral type, the celestial coordinates, $\alpha$ and $\delta$ in
J2000.0, the apparent visual magnitude, $m_{V,\star}$, and the mass,
$M_{\mathrm{S}}$. For the transiting (and more massive) planet the
inputs are the orbital parameters needed for the transit model,
i.e. the semi major axis, $a$, the orbital period, $P$, the
inclination, $i$, the planetary radius, $R_{\mathrm{Trans}}$ and the
mid-transit time, $T_0$, in addition to the planetary mass,
$M_{\mathrm{Trans}}$. For the perturbing planet the inputs are its
mass, $M_{\mathrm{Pert}}$, and the order of the mean-motion resonance,
$j$, since we will consider timing variations caused by an Earth sized
planet in an outer orbit inside a first order resonance
(Section~\ref{sec:Agol}). We then convert the star and both planet
parameters \textbf{into} convenient units for the program. Considering
\mbox{$k \sim 400$ epochs} (this equates to $\sim 2$ years of
follow-up observations), we calculate the transit timing variation
that the perturber exerts on each mid-transit time, $T_{0,k}$.

Instead of fixing the required parameters arbitrarily, we reproduced a
real system: \mbox{Qatar-1} \citep{Alsubai2011}. This is the first
exoplanetary system discovered by the Alsubai Project exoplanet
transit survey. The host has been characterized as an old K~type
star. As a result of the large exoplanetary radius and the short
orbital period of the exoplanet, the transits are deep and easy to
observe, even with small aperture
telescopes. Table~\ref{tab:parameters} shows the orbital parameters
obtained by \cite{Alsubai2011}, considered as input values for our
code. As previously mentioned, our group has been carrying out
follow-up observations of the system for more than two years
\citep[see][]{vonEssen2013}. This allows us to include a comparison
test between real and synthetic data (see Section~\ref{sec:test}).

\begin{table}[t]
  \centering
  \caption[]{Input parameters \citep[Qatar-1,][]{Alsubai2011}. The
    program does not require error estimates. In consequence, they are
    not listed.}
  \label{tab:parameters}
  \begin{tabular}{l l l}
    \hline
    \hline
Star & $M_{\mathrm{S}}$ ($M_{\odot}$) &  0.85   \\
   & $R_{\mathrm{S}}$ ($R_{\odot}$) &  0.823  \\
   & Spectral type     & K2       \\
   & $V$ (mag)         & 12.9    \\
   & $\alpha$ ($^{\mathrm{h}}$) & 20.2251 \\
   & $\delta$ ($^{\circ}$) & 65.1619 \\
    \hline
Planet   & $T_o$ (BJD-TDB)   & 2455518.4102 \\
   & $P$ (days)      & 1.420033  \\
   & i ($^{\circ}$)     & 83.47      \\
   & a ($UA$)          & 0.02343   \\
   & $R_P$ ($R_{\mathrm{J}}$)     & 1.164     \\
   & $M_P$ ($M_{\mathrm{J}}$)     & 1.090     \\
    \hline
Perturber   & $M_{\mathrm{Pert}}$ ($M_{\mathrm{J}}$) & 0.007\\
   & $j$              & 2\\
    \hline
  \end{tabular}
\end{table}

\subsection{Producing the TTV imprint}
\label{sec:Agol}

\citet{Agol2005} derived an order of magnitude of the perturbation
that is caused to the timings of a transiting planet when it coexist
in a first-order mean-motion resonance with a second planet. The
authors estimated the amplitude, \mbox{$\delta t_{\mathrm{max}}$}, and
the libration cycle, $P_{\mathrm{lib}}$, of the timing variations
(their Eq.~33 and Eq.~34, respectively) to be:
\begin{equation}
  \delta t_{\mathrm{max}} \sim \frac{P_{\mathrm{Trans}}}{4.5 j} \frac{m_{\mathrm{Pert}}}{(m_{\mathrm{Pert}} + m_{\mathrm{Trans}})}\; ,
\end{equation}
\begin{equation}
  P_{\mathrm{lib}} \sim 0.5 j^{-4/3} \left(\frac{m_{\mathrm{Trans}}}{m_{\mathrm{Star}}}\right)^{-2/3}\ P_{\mathrm{Trans}}\; .
\end{equation}

\noindent In this work, the perturbations are added to the unperturbed
mid-transit times as follows:

\begin{equation}
  T_{0,k} = T_0 + k \times P_{\mathrm{Trans}} + \delta t_{\mathrm{max}}\ \sin\ [2\pi P_{\mathrm{Trans}} (k -1)/P_{\mathrm{lib}}]\; .
\end{equation}

\noindent $T_0$ is the starting epoch given as input parameter in
Barycentric Julian Dates (\bjdtdb), \mbox{$k \times
  P_{\mathrm{Trans}}$} are the unperturbed mid-transits for each epoch
$k$, and \mbox{$\delta t_{\mathrm{max}}\ \sin\ (2\pi
  P_{\mathrm{Trans}} (k - 1)/P_{\mathrm{lib}})$} is the perturbing
term. Once the TTV signal is added, the program does a main loop over
each perturbed epoch, $T_{0,k}$.

\subsection{The ground-based observatories}
\label{sec:GBO}

For the purposes of our analysis we consider three virtual
observatories. In order to ensure an optimal coverage of observable
transits, they populate the northern hemisphere and are separated
mostly in geographic longitude.

The code requires basic information on the sites and the instrumental
setup, such as the mean seeing, the extinction coefficient, $\kappa$,
the geographic coordinates, the available filters and CCDs, and the
apertures of the primary mirrors. The values considered within our
code are listed in Table~\ref{tab:observatorios}. Particularly, to
perform the photometric follow-up on Qatar-1 we used the telescopes
located at the Hamburger Sternwarte Observatory (HSO) and the
Observatorio Astron\'omico de Mallorca (OAM). Their corresponding
instrumental setup and sky quality descriptions are realistic. In
consequence, the observations collected at both sites will help to
characterize to which extent do synthetic and real data resemble each
other. Although McDonald observatory in reality exists, the
instrumental setup presented here is of our own invention. Although
these parameters are fixed, the program is general and the given
locations, atmospheric characteristics, and equipment sets can be
easily changed to others.

\begin{table*}[t]
  \centering
  \caption[]{Basic description of the observatories considered to
    produce our synthetic light curves. mag/AM denotes magnitudes per
    airmass value. The remaining columns are self-explanatory.}
  \label{tab:observatorios}
  \begin{tabular}{l l l l l c c}
    \hline
    \hline
    Observatory & Geographic coordinates   & Primary mirror (m) & Available
    filters   & CCD     & $\langle$seeing$\rangle$ ($''$)  & $\langle
    \kappa_V \rangle$ (mag/AM)\\
    \hline
    Hamburger   & $\lambda$ = 53.48$^{\circ}$& 1.2           &
    Johnson-Cousins $R$, $I$ & ALTA U 9000  & 2.5       & 0.20 \\
    Sternwarte  & $\phi = 10.2414^{\circ}$ &                & Sloan $i$             &              &               &          \\[0.2cm]
    Observatorio& $\lambda = 39.64^{\circ}$& 0.6           & Johnson-Cousins
    $R$, $I$ &  ST7XM       & 2.0      & 0.18 \\
    Astron\'omico& $\phi = 2.9509^{\circ}$  &               & Sloan $r$             &              &               &          \\
    Mallorca    & & & & & & \\[0.2cm]
    Mc\,Donald   & $\lambda = 30.67^{\circ}$& 1.5           & Sloan $r$, $i$, $z$       & STL11000M    & 1.5      & 0.14 \\
    Observatory & $\phi = -104.021^{\circ}$&                &                     &              &               &          \\
    \hline
  \end{tabular}
\end{table*}

To carry out TTV analysis it is of common use to combine light curves
that were produced under different instrumental setups and
atmospheric conditions \citep[see
  e.g.,][]{Shporer2009,Maciejewski2010,Nascimbeni2013}. To investigate
if this influences the TTV characterization, our program chooses the
observatory and filter randomly.

\subsection{Limb-darkening coefficients}

\cite{ClaretHauschildt2003,Claret2004}, and \cite{Claret2011} provide
the exoplanet community with already calculated limb-darkening
coefficients. Although the authors cover most of (if not all) the
standard systems, many observations are carried out using non-standard
filters. Since it is our intention to make the code as general as
possible, we produced the limb-darkening coefficients in our own
fashion.

As a first step we produced angle-resolved synthetic spectra using
PHOENIX \citep{Peter1,Peter2}, given the effective temperature, the
metallicity, and the surface gravity of the target star.  For Qatar-1,
these values are \mbox{$4861 \pm 125$\,K} for the effective
temperature, \mbox{$\log(g) = 4.536 \pm 0.024$}, and \mbox{[Fe/H]
  $=0.2 \pm 0.1$} \citep{Alsubai2011}. We then convoluted the
synthetic spectra with each filter transmission function (see
Table~\ref{tab:observatorios} for available filters) and CCD quantum
efficiency, and afterwards integrated them in wavelength. We ended up
with intensities as a function of \mbox{$\mu = \cos\ \theta$}, where
$\theta$ is the angle between the line of sight and the line from the
center of the star to a position of the stellar surface. The
normalized intensities are fitted with a quadratic limb-darkening law:

\begin{equation}
  I(\mu)/I(1) = 1 - u_1(1 - \mu) - u_2(1 - \mu)^2\; ,
\end{equation}

\noindent from where the $u_1$ and $u_2$ quadratic limb-darkening
coefficients are computed. The final limb-darkening coefficients are
listed in Table~\ref{tab:limbdarkening}. Once the site, the CCD, and
the filter are randomly chosen, the corresponding limb-darkening
coefficients are added to the program variables.


\begin{table}
  \centering
  \caption{\label{tab:limbdarkening} Quadratic limb-darkening
    coefficients for the filters considered in this simulation.}
  \begin{tabular}{l l l}
    \hline
    \hline
    Filter              &   $u_1$  &   $u_2$  \\
    \hline        
    Johnson-Cousins $R$ &  0.5960  &  0.1147  \\
    Johnson-Cousins $I$ &  0.4669  &  0.1478  \\
    Sloan $r$           &  0.6180  &  0.1086  \\
    Sloan $i$           &  0.4863  &  0.1437  \\
    Sloan $z$           &  0.3812  &  0.1629  \\
    \hline
  \end{tabular}
\end{table}

\subsection{Reference stars}

After the site is selected, the program chooses randomly between one and up to
seven reference stars, which will be later combined to
perform the differential photometry. The selection of the reference
stars ({\it rs}) complies one of the following three criteria:

\begin{itemize}
\item The {\it rs} are the same for all the sites along all
  epochs. Therefore, the program will choose them only once.
\item The {\it rs} are the same, but for each site only. Therefore,
  three different sets of reference stars will be chosen once.
\item The {\it rs} will be always different. Therefore, the number of
  {\it rs}, their angular separation relative to the target star in
  ($\Delta \alpha$,$\Delta \delta$), and their spectral type, will be
  selected during each epoch.
\end{itemize}

The celestial coordinates of the target star are precessed from
J2000.0 to $T_{0,k}$. Then the \mbox{($\Delta \alpha$,$\Delta
  \delta$)} separations of the {\it rs}, relative to the target star,
are randomly determined. The maximum values that \mbox{$\Delta
  \alpha$} and \mbox{$\Delta \delta$} can take are limited by the
telescope's field of view, while the minimum values are set to five
times the mean seeing of the site. In a further step, another
subroutine assigns the spectral and sub-spectral type, from where the
effective temperatures, $T_{\mathrm{eff}}$, are added to the program
variables.

Instead of assigning the spectral type to the {\it rs} from a flat
distribution (i.e., any spectral type has the same probability to be
randomly selected), we carried out a more realistic approach. To this
end, we used an extended version of the Henry Draper (HD) catalog
\citep[HDEC,][]{Nesterov1995,harchenko2009}. The catalog provides,
among others, the spectral types of $\sim 88\,000$ stars as a function
of their apparent magnitude. Is this what it makes it so suitable for
our purposes: rather than using the true stellar spectral type
distribution, the catalog represents more realistically the
distribution of observable stars in a given magnitude range. Since
accurate photometric light curves are generally obtained when the
brightness difference between target and {\it rs} is small \citep[see
  e.g.,][]{Howell2006}, the knowledge of the apparent magnitude of the
target star would set constraints on the fluxes of convenient {\it
  rs}. One adequate limit, considered within our code, is given by
\mbox{$\Delta m = | m_{V,\star} - m_{V,\mathrm{ref}}| <
  0.5$}. Therefore, we firstly selected from the catalog the stars
within the range of magnitudes $(m_{V,\star}-0.5,\; m_{V,\star}+0.5)$,
where $(m_{V,\star}$ and $m_{V,\mathrm{ref}}$ correspond to the visual
magnitude of Qatar-1 and the field stars, respectively. Then, we
counted the number of stars per spectral types, O,B,A,F,G,K, and
M. The resulting histogram is listed in Table~\ref{tab:ST}. Finally,
to assign the spectral types to the reference stars, we used the
normalized histogram as probability distribution function. Although
the HD catalog is quite extensive, the number of stars is relatively
low to produce a further discrimination with respect to the
sub-spectral type. Therefore, the sub-class is randomly chosen.

\begin{table}[ht!]
  \centering
  \caption{\label{tab:ST}Stellar number as a function of spectral type
    for the stars with magnitudes close to $m_{V,\star}$ (14\,077 out of
    $\sim88\,000$ stars) obtained from the Henry Draper catalog.}
  \begin{tabular}{c c}
    \hline
    \hline
    Spectral type & Number of stars \\
    \hline
    O & 21\\
    B & 511\\
    A & 5058\\
    F & 2648\\
    G & 2542\\
    K & 2327\\
    M & 970\\
    \hline
  \end{tabular}
\end{table}

\subsection{Visibility}

In a further step, the program verifies whether the mid-transit time,
$T_{0,k}$, occurs at night. As ``night'', we consider the time between
astronomical twilights. Therefore, we calculate sunrise and sunset
times for a Sun at $-18^{\circ}$ with respect to the selected site's
horizon. For this, we make use of the celestial coordinates of the
target star and the geographic coordinates of the observatories, along
with their altitude above sea level. If the transit occurs during
night, we inspect whether the star's altitude at mid transit is higher
than $35^{\circ}$. This is rendered to avoid non-linear extinction
effects in our synthetic light curves. If, however, $T_{0,k}$ occurs
during daylight or during night, but with the star under $35^{\circ}$
of altitude, then the program skips the rest, increments one epoch,
and repeats all the steps again up to this one.

\subsection{Duration of the observations}

With the mid-transit time taking place at night, and the star above
35$^{\circ}$, the program produces a random length for the synthetic
transit light curves. As time scale we use the transit duration
$T_{\mathrm{dur}}$, which can be estimated from the system's orbital
parameters \citep{Haswell2010}:
\begin{equation}
  T_{\mathrm{dur}} = \frac{P}{\pi} \textrm{asin}\left(\frac{\sqrt{(R_{\mathrm{S}} + R_{\mathrm{P}})^2 - a \cos(i)^2}}{a}\right)\; .
\end{equation}
In order to ensure synthetic light curves as realistic as
possible, with the calculated observation length the program randomly
selects one among four scenarios:

\begin{itemize}
\item The transit is complete, including also a considerable amount of
  out of transit (OOT) data before and after transit.
\item The transit is partially complete. It includes OOT data
  before and after transit, but has also data gaps in between.
\item The transit is not complete. The mid-transit time is observed
  but ingress or egress and OOT data before or after transit,
  respectively, are completely missing.
\item The transit is not complete. The mid-transit time is not
  observed. Only a small fraction of ingress or egress along with some
  OOT data are produced.
\end{itemize}

The segment of the transit that is missing and the longitude of the
OOT data points are always a random multiple of the transit
duration. Particularly, the duration of the OOT data points is smaller
than 2-3 hours, since real observations of primary transit events tend
to be produced in this fashion.

Subsequently, the program estimates the exposure time, considered as
fixed within each epoch. This resembles observations performed by
robotic telescopes, for which the exposure time is estimated a priory
in order to reach certain signal-to-noise ratio. The exposure time is
estimated considering the telescope's primary mirror size, the
altitude of the star at mid-transit time, the star visual magnitude,
the filter response, the atmospheric mean extinction, the CCD quantum
efficiency, the Moon phase estimated for each epoch, and the desired
signal-to-noise ratio.

\subsection{Time stamps}

Making use of the duration of the observation and the exposure time as
time steps, we produce a temporal array in universal time
(equivalently, in Julian dates). Using \citet{Eastman2010} web
tool\footnote[1]{http://astroutils.astronomy.ohio-state.edu/time/utc2bjd.html},
we then convert the Julian dates into barycentric Julian dates
employing the celestial coordinates of the star in J2000.0, the
geographic coordinates of the site, and its height above sea
level. With the time stamps in barycentric Julian dates, we calculate
the projected separation between planet and star centers, $\delta_j$,
for each instant $BJD_j$:

\begin{equation}
  \delta_j = \sqrt{1 - \cos(\phi_j)^2 \sin(i)^2}~\frac{a}{R_{\mathrm{S}}}\; ,
\end{equation}

\noindent which requires the previous knowledge of the orbital phase:

\begin{equation}
  \phi_j = \frac{2\pi~(BJD_j - T_{0,k})}{P} - n_{\mathrm{orb}}\; ,
\end{equation}

\noindent for a given orbit number $n_{\mathrm{orb}}$. Using
$\delta_j$, the planet-to-star radius ratio, $p =
R_{\mathrm{P}}/R_{\mathrm{S}}$, the impact parameter, $a\cos(i)$, and
the quadratic limb-darkening coefficients, $u_1$ and $u_2$, we produce
the synthetic star flux-drop during transit using the primary transit
model provided by \citet{MandelAgol2002}. Once the basic structure of
the light curves is complete (i.e., primary transit model as a
function of barycentric Julian dates) we add noise associated to our
Earth's atmosphere, to the instrumental configuration, and to the
intrinsic variability of the host star, saving the product at each
step.


\section{Our code: Correlated noise sources}


In general, the unwanted variations in the observations that are not
due to random (i.e., photon) noise are correlated in time.  When it
comes to ground-based observations, this variability is mainly caused
by transparency variations in the Earth's atmosphere, changes in the
altitude (i.e., airmass) of the star along the observations,
imperfections in the instrumentation used to carry out the
observations, and stellar variability that a given photometric
precision does not allow to properly account for. Therefore, the
natural scatter in the data will not be white but ``red''
instead. ``Red noise'' is the manifestation of systematic effects in
photometric time series, and is ``red''-colored because it has
low-frequency (i.e., time-correlated) components; in transit
observations it manifests itself in the milli-magnitude regime. It was
\cite{Pont2006} who first raised the importance of red noise in
exoplanet time-series. The main goal of this work is to model as
realistically as possible noise structure that is typically present in
ground-based data, and to study to which extent are the mid-transit
times affected by it.


\subsection{Intrinsic variability: occulted and unocculted spots}
\label{sec:bc}

Owing to the high photometric quality provided by space-based
observations such as CoRoT \citep{Auvergne2009} and Kepler
\citep{Borucki2010,Koch2010}, stellar magnetic activity and its impact
over transit light curves has been studied in great detail. Dark spots
and bright faculae on the stellar photosphere move as the star
rotates, producing a time-dependent variation of the stellar flux,
which can have a significant impact on the computation of planetary
and stellar parameters \citep[see e.g.,][]{Czesla2009,Lanza2011}.

Occulted and unocculted spots can affect the shape of transit light
curves. Current achievable photometric precision allows us to use the
small imprints of occulted spots on transit data to characterize the
stellar surface brightness profile and the spot migration and
evolution \citep[see
  e.g.,][]{Carter2011,Sanchis-Ojeda2011a,Sanchis-Ojeda2011b,Huber2010}.
When transit fitting is performed, an incorrect treatment of
unocculted spots can lead to an incorrect characterization of transit
parameters, already observed when the wavelength-dependent variability
of the transit depth is being reviewed \citep[see e.g.,][when light
  curves with quality like the ones produced here are being used to
  characterize the exoplanet atmosphere]{Mallonn2015}. Since the
stellar surface is continuously evolving, the impact of unocculted
spots also change from transit to transit. In the context of TTVs,
several studies have already been carried out to characterize the
impact of spots over the determination of mid-transit times. Actually,
the deformations that stellar activity produces over the transits have
been studied in detail and pinpointed in some cases as a misleading
identification for TTVs \citep[see
  e.g.,][]{Rabus2009,Maciejewski2011,Barros2013}. A recent example
involving simulations is provided by \cite{Ioannidis2016}. Among
others, when relatively low signal-to-noise transit light curves were
analyzed (such as the ones produced in this work) the authors point
out the difficulty to determine whether artificially injected TTV
periods can be identified as due to starspots, physical companions or
random noise artifacts.

To simulate the effect caused by occulted (dark) spots over our
transit light curves we first characterized their expected
amplitude. Since we are only interested in an order of magnitude, our
simulations are carried out considering a star whose center-to-limb
variability is represented by a quadratic limb darkening law. Thus, to
be consistent we used the limb darkening coefficients specified in
Table~\ref{tab:limbdarkening}. We also considered only one spot in the
center of the star, with sizes set to one-third, one-fourth and
one-fifth the size of the transiting exoplanet. To compute the
contrast between the stellar photosphere and the spot, we treated both
star and spot as black bodies. The effective temperature of
\mbox{Qatar-1} is the one determined by \cite{Alsubai2011}. To
estimate the spots effective temperatures we used the spot temperature
contrast data observed in Figure 5 of \cite{Andersen2015}. Around
\mbox{Qatar-1's} effective temperature, the expected spot temperatures
show a large scatter. Therefore, for our simulations we have
considered spot effective temperatures which have a difference with
Qatar-1's photosphere of 700 to 1300 K, and considered a step of 100
K. We convoluted both black bodies with the photometric filters
specified in Table~\ref{tab:limbdarkening}, and we computed the ratio
of the derived spot and star fluxes. This flux ratio was used to set
the level relative to the photosphere of the synthetic spot. We
finally crossed a planet with orbital and physical parameters matching
the ones of \mbox{Qatar-1 b}, and computed the amplitude of the
``bump'' from the difference between a transit with an occulted and an
unocculted spot. As expected, our results show a clear dependency with
wavelength, temperature difference, and spot size. A typical bump
amplitude was found to be between 2 and 12\% the transit depth. In our
code, we randomly select three spot temperatures and three
planet-to-spot size ratios from the values previously specified. We
then assign to these their corresponding bump amplitude depending on
the filter. Although we choose three set of parameters we do not
produce three bumps in the light curves. The bump number and position
is assigned also randomly, being the possible number between 0 and 2,
which is what observations commonly show \citep[see
  e.g.,][]{Sanchis-Ojeda2011b,Sanchis-Ojeda2012}.

Unocculted spots change the overall level of the light curve,
producing a time-dependent modulation. Indeed, \cite{Czesla2009}
already observed a detectable variation in the transit depth when
light curves with the highest and lowest continuum levels are being
compared (see their Fig.~2). The amplitude of the variability depends
on the activity level, the spot coverage, and the contrast between the
spot and stellar photosphere temperatures. However, the variability
they observe in the transits of CoRoT-2, one of the most active planet
host stars \citep[see][and references therein]{Huber2010}, is within a
few percent of the planet-to-star radius ratio. Further examples of
the impact of unocculted spots can be observed when the exoplanet
transmission spectrum is being retrieved \citep[see
  e.g.,][]{Sing2011,Mancini2014}. To account for the effect of
unocculted spots we begin by simulating the overall variability on the
light curve. Spots move with the star as it rotates. As rotation
period we use the value obtained by \cite{Mislis2015}, $P_1 = 23.697$
days. To consider differential rotation, as well as the fact that
spots might change their location, size, or occurrence, we randomly
choose a second period, $P_2$, that should follow the relation given
by \cite{Reinhold2015}:

\begin{equation}
0.01 \leq | P_1 - P_2 | \leq 0.30\;.
\end{equation}

\noindent Once the rotation periods are selected, we randomly select
two phases, $\phi_1$ and $\phi_2$, from a uniform distribution between
0 and 1. Finally, the semi-amplitudes of the spot modulations, $A_1$
and $A_2$ (satisfying $A_1 > A_2$), are chosen from a uniform
distribution between 0 and 20 parts per thousand. This upper limit
corresponds to the semi-amplitude of the spot modulation (SM) observed
in \mbox{CoRoT-2}. The SM is then represented as follows:

\begin{equation}
 SM(t) = A_1 \sin[2\pi(t/P_1 + \phi_1)] + A_2 \sin[2\pi(t/P_2 + \phi_2)]\;,
\end{equation}

\noindent evaluated during the time ($t$) of the observation. After
adding the spot modulation to the light curves, the next step is to
consider how stellar activity affects the derived size of extrasolar
planets. As maximum amplitude variability of
\mbox{$R_{\mathrm{P}}/R_{\mathrm{S}}$} we used the 3\% value derived
from \cite{Czesla2009}, but modulated by the activity phase of the
star. In other words, if the transit occurs close to a maximum flux of
the star (minimum spot coverage) then the amplitude will be close to
0\%. If, however, the transit occurs during a minimum of stellar flux,
the amplitude variability of \mbox{$R_{\mathrm{P}}/R_{\mathrm{S}}$}
will be close to the largest possible value. The variability is set
around the mean value of \mbox{$R_{\mathrm{P}}/R_{\mathrm{S}}$};
therefore, \mbox{$R_{\mathrm{P}}/R_{\mathrm{S}}$} will vary between
\mbox{$\pm 1.5$\%}.

Since we want to study the impact of stellar activity into the
determination of TTVs when ground-based data are being analyzed, we
carried out our simulations with and without accounting for stellar
activity. However, the impact of spots in the context of ground-based
TTV characterization deserves a deeper discussion than the one we can
provide here. Therefore, we will present these results in another
publication. For the simulations carried out in this work, spots have
been shut down.

\subsection{Instrumental systematics}
\label{sec:d}

\begin{figure}[ht!]
  \centering
  \includegraphics[height=.0908\textheight]{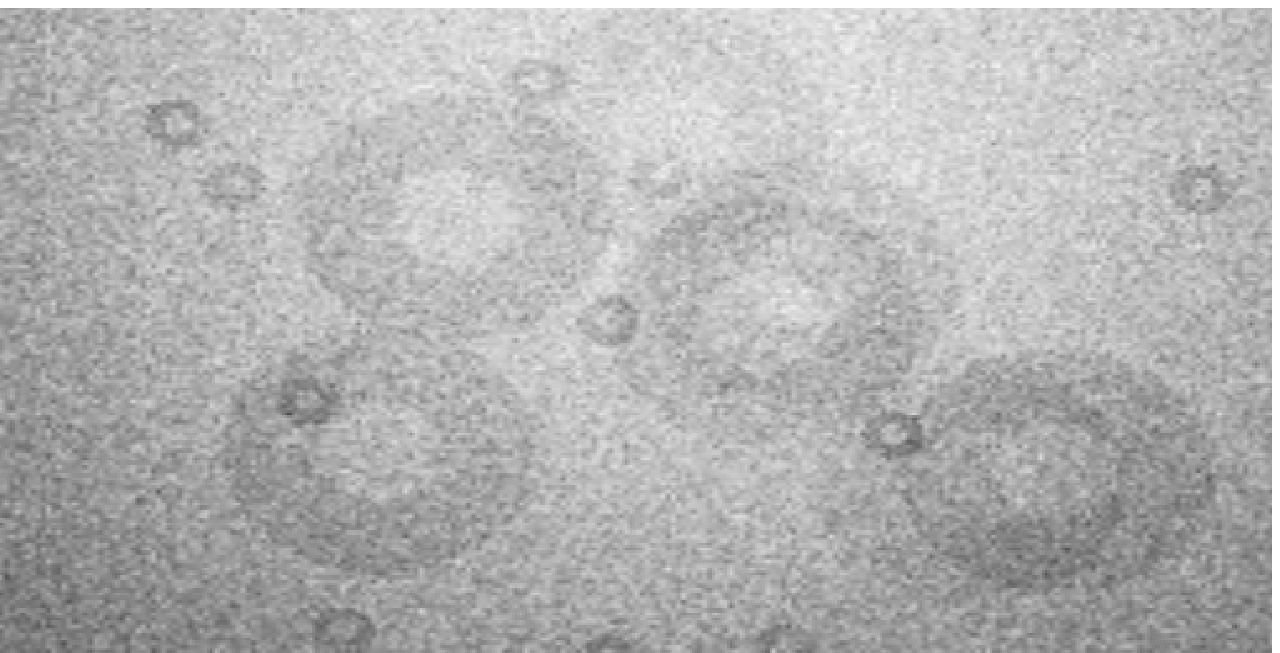}%
  \includegraphics[height=.0908\textheight]{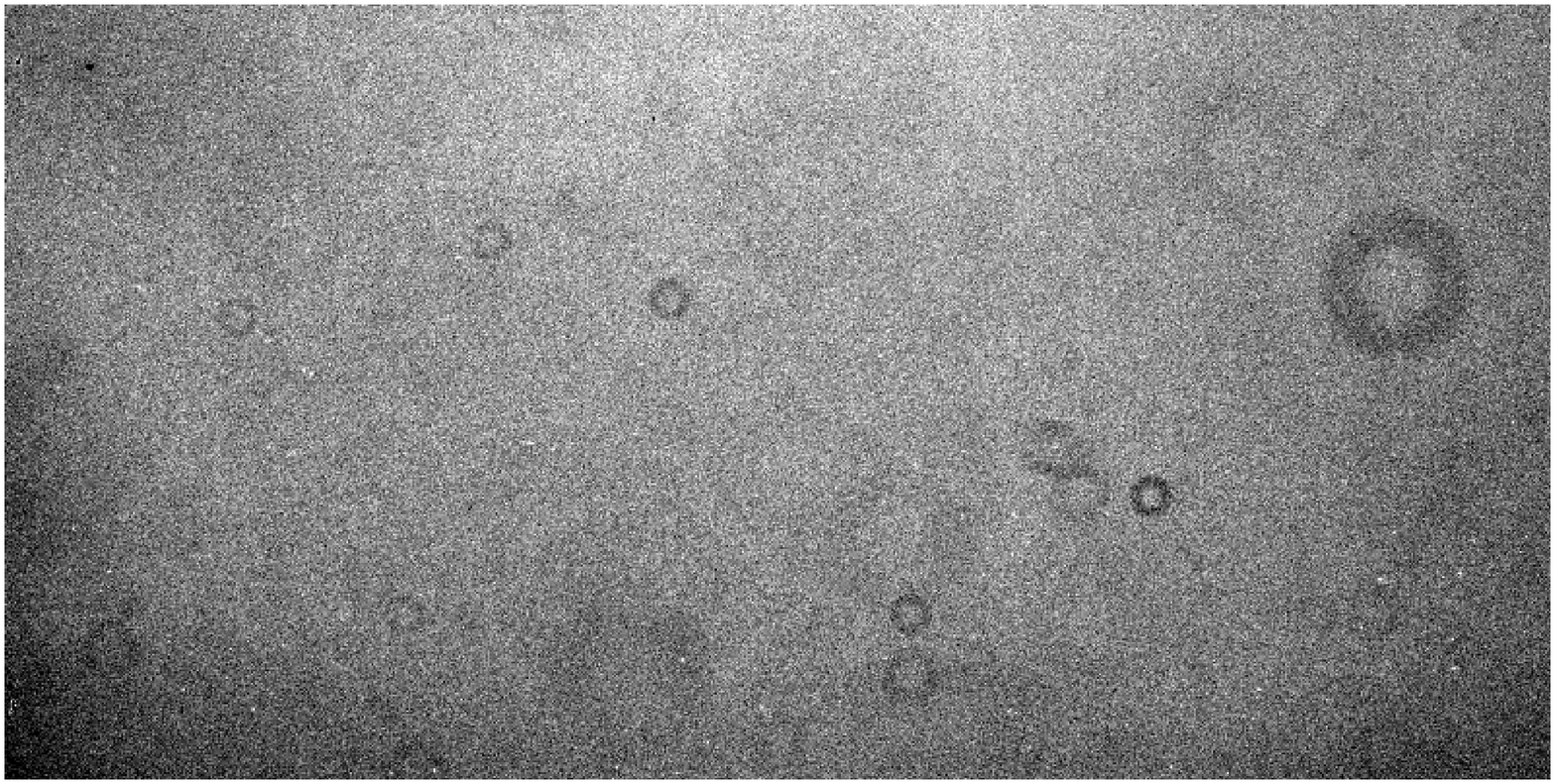}

  \includegraphics[width=.485\textwidth]{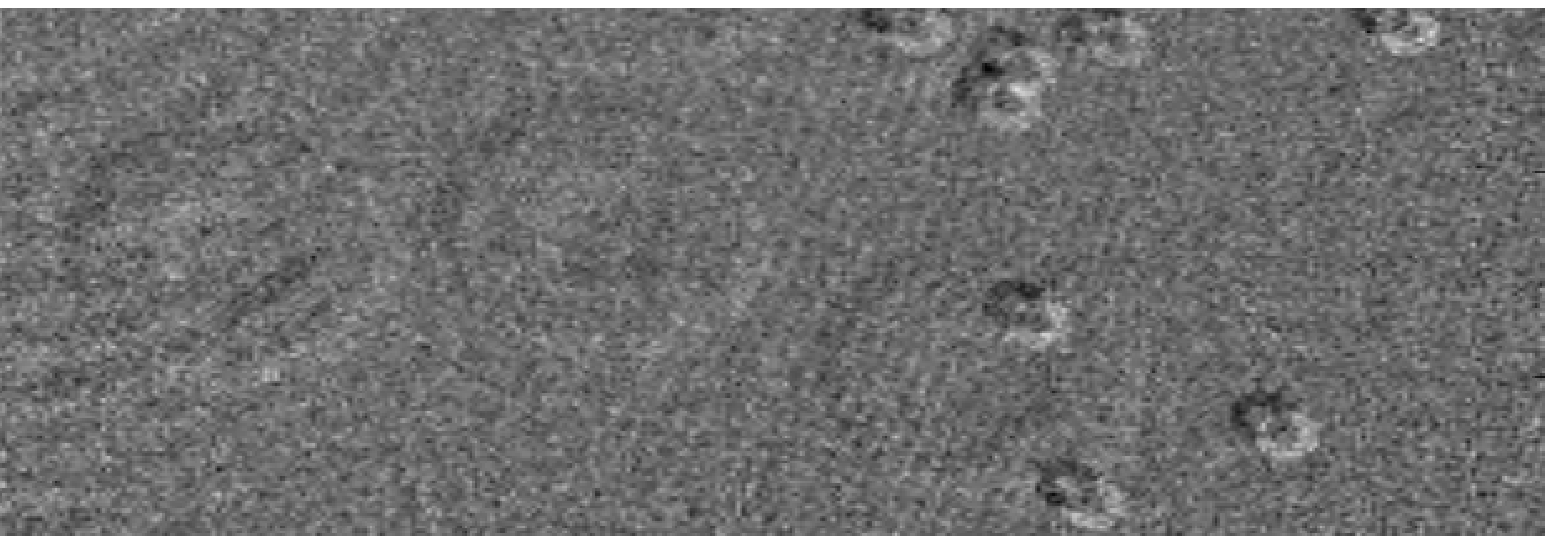}
  \caption{\label{fig:flats}Fraction of simulated flat fields. {\it
      Top, left:} two sized dust grains, pixel-to-pixel sensitivity
    variability and uneven illumination can be observed. {\it Top,
      right:} as comparison, a real flat frame. The pixel and gray
    scales are different. {\it Bottom:} how a science frame would look
    like when taken during Moon light. The dark and bright spots can
    be easily observed.}
\end{figure}

Flatfield frames are obtained during photometric runs to remove mainly
pixel-to-pixel sensitivity variations and (typically two-sized)
defocused images of dust grains sitting on the filter wheel and on the
CCD. Usually, when the telescope is not defocused \citep[see e.g.,
][]{Kjeldsen1992,Southworth2009} or when it is not guiding, once
science frames are bias-subtracted and flatfielded, small
imperfections can be observed. This can be acknowledged by visually
inspecting the calibrated science frames, and sometimes by observing
that the red noise in the light curves is correlated with the centroid
position of the stars, which might change in time due to drifts on the
detector caused from imperfect tracking and/or seeing variations. This
residual variability is mostly caused by the finite precision that the
flat fields have to reproduce the imperfections on the CCD. In this
work we have simulated flats reproducing some features that can be
usually observed in such frames: pixel-to-pixel variability, central
excess illumination, cosmetics caused during the construction of the
CCD, and shadows of dust grains, which can take two different sizes
depending on if they are located directly over the CCD (small size) or
over the filters (large size). Since most of these parameters depend
on the CCD quality, the values that the flat field take in our
simulations depend on the selected observatory. We have also added an
effect that is caused by Moonlight, thus correlated with the Moon
phase: when skyflats or domeflats are obtained, the illumination is
expected to be homogeneous. However, when observations are carried out
under Moonlight, the shadow of the dust grains is projected in a
slightly different direction, depending on where the Moon is in the
sky relative to the target star during the observations. Therefore,
when flat fields are used to correct science frames obtained during
gray/bright nights, bright and dark spots can be observed sometimes
even by eye in the location of the dust grains. This effect, along
with the synthetic flat fields produced in this work, can be observed
in Figure~\ref{fig:flats}.

In this work we include the flat field residual modulation as follows:
the code randomly chooses two positions over the synthetic flat field
representing the positions of the target and one reference star. The
distance between both is chosen randomly. Then, their positions are
drifted over the image, creating an (x,y)($t$) array whose length is
equal to the length of a given observation. Depending of the telescope
in question, the amplitude of the drift is going to be larger (OAM,
assuming no guiding system) or smaller (HSO and MCD, assuming some
sort of guiding). Also, we consider the (x,y) drift to be more
prominent in one direction, assuming that the telescope is drifting
more in right ascention, as it is normally the case with real
telescopes. The quality of the flat (the pixel-to-pixel variability)
also depends on the telescope. We assumed that the larger the
telescope, the more expensive the CCD is and, therefore, the more
accurate the flat is. We then integrate flat counts inside an aperture
that corresponds to an integration area of 40 pixels$^2$, and divide
the count rate of the target by the count rate of the reference star
in each time stamp. Finally, we normalize and scale the amplitude of
the modulation down to a random number between 2 and 10
parts-per-thousand, to meet typical amplitudes of instrumental
modulation. The computed variability is then saved, along with the
(x,y) position, to be used during the detrending instance.

\subsection{Residual modulation due to first order atmospheric extinction}
\label{sec:e}

First order atmospheric extinction, (i.e., extinction independent of
stellar color), is airmass dependent. Since differential photometry
involves at least two stars at different elevations, a residual
modulation due to airmass differences can be detected, increasing when
the elevation difference between the target star (sub-index $\star$) and the
reference stars (sub-index $1, 2, \cdots, n$) increases as well. For
any star, absorption by the atmosphere can be described by Bouguer's
law:
\begin{equation}
  m = m_0 - \kappa \chi\; ,
\end{equation}
where $m_0$ denotes the stellar magnitude outside the atmosphere,
$\kappa$ the extinction coefficient in magnitudes per airmass
(mag/AM), $\chi = \sec(z)$ the airmass value during a certain
observation, and $z$ the zenithal distance of the star. Since light
curves are produced only when the altitude of the star at mid-transit
time is larger than 35$^{\circ}$, the linear representation of airmass
is sufficiently accurate.

To decrease the scatter of the final light curve, it is of common
practice to consider as reference star the combination of many
others. Hence, considering Bouguer's and Pogson's laws, the airmass
modulation, $AM_{\mathrm{mod}}$, that will affect the differential light curve
will follow:
\begin{equation}
  \log(AM_{\mathrm{mod}}) = \frac{\kappa}{2.5}\left(\chi_\star - \frac{1}{n}\sum_{i=1}^n\chi_i\right)\; .
  \label{eq:AM_MOD}
\end{equation}
The second term between the parentheses accounts for the
combined airmass contribution of $n$ reference stars.

\begin{figure}[ht!]
  \centering
  \includegraphics[width=.5\textwidth]{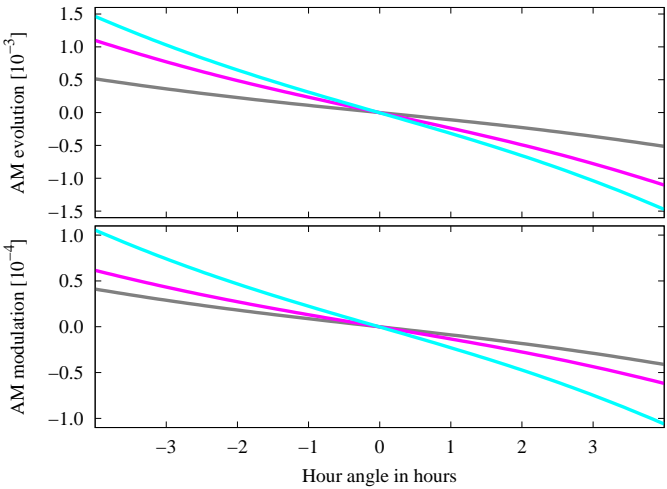}
  \caption{\label{fig:airmasscorr} {\it Top:} airmass difference
    between target and reference star, as a function of hour angle, in
    units of $10^{-3}$. {\it Bottom:} airmass modulation due to
    airmass differences (Eq.~\ref{eq:AM_MOD}), in units of
    $10^{-4}$. The lines correspond to an angular separation of 7
    arcmin and \mbox{$\kappa_{HSO}$ = 0.2 mag/AM} (gray), 15 arcmin
    and \mbox{$\kappa_{MCD}$ = 0.14 mag/AM} (pink) and 20 arcmin and
    \mbox{$\kappa_{OAM}$ = 0.18 mag/AM} (cyan).}
\end{figure}

The top panel of Figure~\ref{fig:airmasscorr} shows how the airmass
difference between target and one particular reference star evolves,
as a function of the angular separation between stellar objects
(color-coded) and the hour angle, $t$, where \mbox{$t = 0$} denotes the
culmination of the target star. The bottom panel of the same Figure
shows how the airmass modulation $AM_{\mathrm{mod}}$ evolves, as the stars move
across the sky. Note that the angular separation between stars is
constrained by the size of the field of view of each telescope.

To calculate the airmass, $\chi_{i}$, for each reference star, we use
the previously selected $\Delta \alpha$ and $\Delta \delta$
displacements, relative to the target star. Although the modulation
effect is small, it rapidly increases with the angular separation
between target and reference stars. Since small telescopes tend to
have large fields of view ($\sim 1$ deg or larger) we consider this
effect relevant and the first atmospheric-correlated noise source. The
unaffected transit light curve is deformed by $AM_{\mathrm{mod}}$.


\subsection{Color-dependent residual modulation}
\label{sec:f}

Extinction is caused by absorption and scattering of light. Water
vapor, ozone, and dust, but mostly Rayleigh scattering in the optical,
are contributing to it. Color-dependent extinction (or ``second-order
extinction'') appears because the light of a stellar object, on its
path through the atmosphere, has a wavelength-dependent absorption. In
consequence, if two stars of dissimilar intrinsic color indexes are
observed at the same altitude, their respective absorption will differ.

\begin{figure}[ht!]
  \centering
  \includegraphics[width=.5\textwidth]{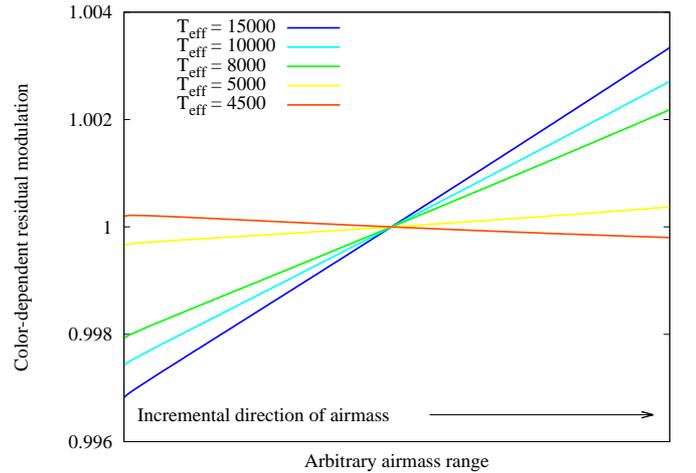}
  \caption{\label{fig:2nd_order_ext}Color-dependent residual
    modulation considering different effective temperatures for the
    reference stars.}
\end{figure}

When the differential photometry technique is performed using stars of
different spectral types, a color-dependent residual shows
up. Frequently, the spectral information of the stars involved in the
differential photometry is completely missing. Therefore, the effect
can not be modeled out. To model it in, we followed \cite{Broeg2005}
methodology (see their Section 4.2 for a complete analytic
description). For each observation instant $j$, the second order
extinction modulation $SOE_{\mathrm{Mod}}$ follows:
\begin{equation}
  SOE_{\mathrm{Mod}} = R(T_\star, \chi_{\star,j})/(\Pi_{i=1}^n\ R(T_i, \chi_{i,j}))^{1/n}\; ,
\end{equation}
 where \mbox{$R(T_\star, \chi_{\star,j})$} accounts for the flux
 change of the target star due to the Earth's atmosphere, while
 \mbox{$R(T_i, \chi_{i,j})$} does so for each one of the reference
 stars $i = 1, \ldots, n$. The wavelength dependency has been already
 integrated out. It involves the filter transmission function, the
 quantum efficiency of the CCD, and the black-body curves of the
 target and reference stars. Figure~\ref{fig:2nd_order_ext} shows how
 the second order extinction amplitude depends on the spectral type of
 the chosen reference stars. Considering a target star with
 $T_{\mathrm{eff}}$ = 4\,900\,K (similar to Qatar-1), we estimated the
 strength of $SOE_{\mathrm{Mod}}$ for one given reference star with
 effective temperatures 3\,000, 4\,500, 5\,000, 8\,000, 10\,000, and
 15\,000 K. As the Figure clearly reveals, the effect grows when the
 difference between spectral types maximizes. Note that the slope of
 the residual modulation changes from positive to negative, when the
 effective temperature of the reference star turns from being larger
 to smaller than the effective temperature of the target star.

\subsection{Scintillation}
\label{sec:g}

Astronomical seeing refers to the blurring of astronomical images
caused by the turbulence in the Earth's atmosphere. In addition, the
brightness of stars appears to vary due to scintillation, which is
caused by small-scale fluctuations in the air density as a result of
temperature gradients. Based on \cite{Young1993}'s approach, we
estimate the contribution of scintillation noise to the accuracy of
photometric measurements:

\begin{equation}
  S = 0.0030\ D^{-2/3}\ \chi^{3/2}\ e^{-h_{\mathrm{obs}}/h_0}\ \tau^{-1/2}\; ,
\end{equation}

\noindent where $D$ is the telescope diameter in meters,
$h_{\mathrm{obs}}$ is the altitude of the observatory above sea level
in km for $h_0$ = 8 km, and $\tau$ the exposure time in minutes. In
differential light curves scintillation translates directly into the
scatter of the data. To include this effect into our light curves, we
calculate random Gaussian noise with $\mu = 1$ and $\sigma$ equal to
the given scintillation semi-amplitude. The only changing factor on
Young's scintillation expression is the airmass, so the standard
deviation will not be constant but will be modulated by the star's
altitude, as seen in real light curves, where photometric precision
decreases with $\chi$ for a fixed exposure time. Consequently, the
primary transit synthetic light curves account for scintillation as
well.

\subsection{Non-photometric conditions}
\label{sec:npc}

\subsubsection{Irregularities caused by changes in the atmospheric seeing}
\label{sec:npc_seeing}

Although large telescopes are located in the most convenient sites
with respect to altitude and photometric conditions, this is not
always the case for decimeter--to--meter class telescopes. Small
telescopes are located all over the world, where photometric
conditions can be far from optimal. Abrupt changes in the atmospheric
transparency, the humidity and the ambient temperature, added to
cirrus and clouds passing by, can produce unwanted photometric
variability. In such sites, atmospheric seeing tends to quickly
degrade with airmass.

Aperture photometry involves the measurement of stellar fluxes within
a fixed aperture radius. Thus, during any data reduction process the
aperture radius can be selected to coincide with, for example, the
full-width at half-maximum (FWHM) of the first image. For instance, if
the observations are carried out only after culmination, as a product
of the degradation of the atmospheric seeing the integrated flux
inside the fixed aperture will decrease with time. If changes in the
photometric conditions would propagate equally to all the stars within
the field of view, the differential photometry technique would be
satisfactory to remove correlated noise produced by those changes,
although the amount of white noise would still change as a function of
time, since the number of photons collected in each aperture would
change as the stars cross the sky. Nonetheless, real photometry
reveals that the point spread function (PSF) of all the stars slightly
differ from one another. Therefore, differential light curves will
show a residual modulation strongly correlated with airmass. To model
this effect, we made use of physically and empirically motivated
relationships. 

\begin{figure}[ht!]
  \centering
  \includegraphics[width=.5\textwidth]{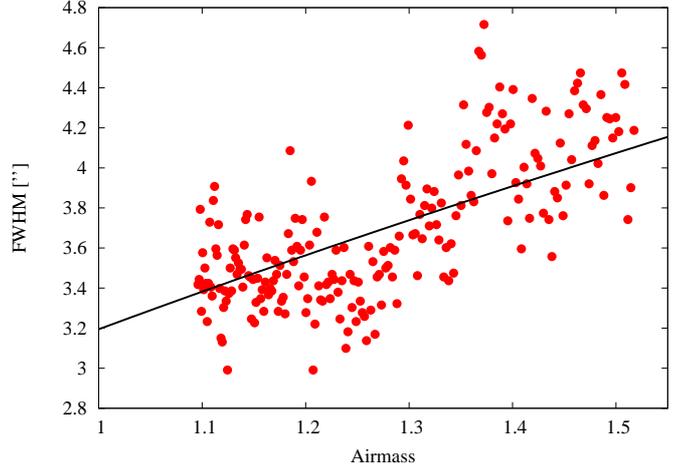}

  \caption{\label{fig:seeing_vs_airmass}FWHM evolution as a function
    of the airmass. Red points show true measurements of FWHM
    (arcsec), while black continuous line indicates the
    \mbox{$FWHM(\chi) \propto \chi^{0.6}$} dependency.}
\end{figure}

Although atmospheric seeing is a very local measurement that strongly
depends on the position of the turbulent atmospheric layers, we
started considering seeing as scaling with the airmass to the power of
0.6 \citep[see
  e.g.,][]{Sarazin1990,Gusev2011}. Figure~\ref{fig:seeing_vs_airmass}
shows the evolution of seeing (equivalently, FWHM) as a function of
airmass. The FWHM measurements correspond to Qatar-1, from
observations carried out at Hamburger Sternwarte. The black continuous
line indicates a fit to the data of the form \mbox{$FWHM(\chi) \propto
  \chi^{0.6}$}. As it can be seen, the relation properly reproduces
the FWHM general trend. During the observations the telescope was
slightly defocused. Therefore, the values of the FWHM are not a
realistic measurement of the characteristic seeing of the site.

\begin{figure}[ht!]
  \centering
  \includegraphics[width=.5\textwidth]{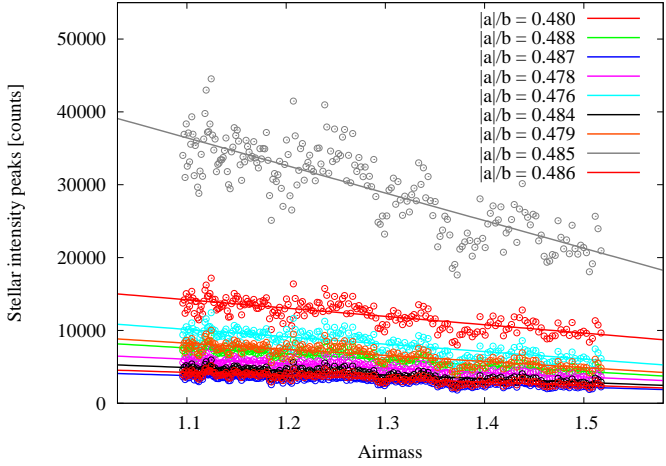}
  \caption{Intensity peaks of 9 stars within the field of view of HSO,
    as a function of airmass (saturation level of HSO CCD lies at
    65535 counts). Linear trends, as reported in Eq.~\ref{eq:AmpPeak},
    are over-plotted in continuous
    lines. \label{fig:amplitude_airmass}}
\end{figure}

Furthermore, the stellar integrated fluxes are estimated as the area
within a two-dimensional normal function, \mbox{$G(\mu,\sigma)$}. The
FWHM is related to the normal function via the standard deviation,
$\sigma$, as \mbox{$FWHM = 2\sqrt{2 \ln(2)}~\sigma$}. For a given
aperture radius $R_{\mathrm{ap}}$ and FWHM, the area is easy to
integrate. In polar coordinates, for any given star:

\begin{eqnarray}
  F_{\star} &=& \int_0^{2\pi}\int_0^{R_{ap}} G(r,\theta)~ r~ dr~ d\theta \nonumber \\
         &=& \int_0^{2\pi}\int_0^{R_{ap}} A\ e^{-r^2/2\sigma^2}~ r~ dr~ d\theta\;,
\end{eqnarray}

\noindent where $A$ denotes the intensity peak of the normal
function. For a fixed exposure time, an increase in airmass translates
into a decrease in the intensity peak. To shape this out, we studied
the intensity peak evolution present in our observations of
Qatar-1. Figure~\ref{fig:amplitude_airmass} shows the evolution of $A$
as a function of airmass, for \mbox{$n = 9$} stars. From our combined
HSO and OAM data we found that a linear relation, in the form:

\begin{equation}
  A(\chi)_i = -a_i\chi + b_i\; , a > 0\;,
  \label{eq:AmpPeak}
\end{equation}

\noindent is sufficient to properly reproduce the observed
variation. Furthermore, the relation between the slope and the
intercept satisfies:
\begin{equation}
  |a_i|/b_i =  \mathbb{C} + \epsilon_i\;,
  \label{eq:AB}
\end{equation}
 for each star $i$ within the field of view, for
\mbox{$\epsilon \ll 0$}, and $\mathbb{C}$ a number close to
0.5. Independently of the intrinsic brightness of the stars, our
observations reveal that the ratio \mbox{$|a_i|/b_i$} remains
approximately the same during a given observing run, as reflected in
Eq.~\ref{eq:AB}.

\begin{figure}[ht!]
  \centering
  \includegraphics[width=.5\textwidth]{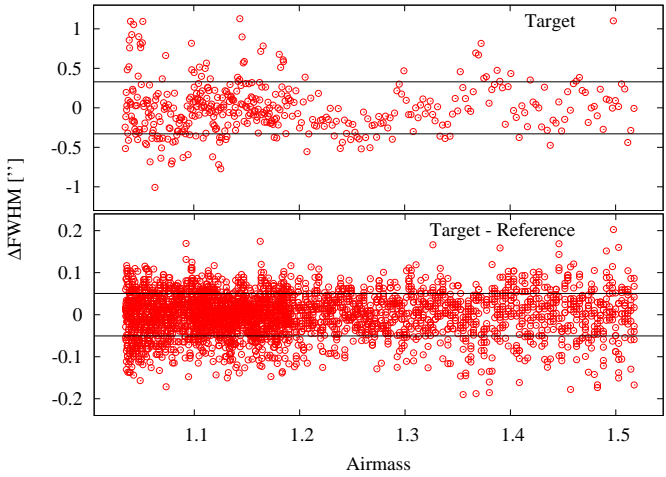}

  \caption{\label{fig:DeltaFWHM}Changes in the FWHM for Qatar-1 (top)
    and for eight reference stars (bottom) relative to
    Qatar-1. Airmass values between 1 and 1.2 are over-sampled with
    respect to the rest, because the star was observed before and
    after culmination. The standard distribution of both data sets was
    used to set constraints on the scatter of the simulated FWHM.}
\end{figure}

With the FWHM and $A$ empirically described as a function of airmass,
we re-analyzed our observations to set constraints on the dispersion
of both parameters. As an example, the top panel of
Figure~\ref{fig:DeltaFWHM} shows the variation of the FWHM for Qatar-1
with respect to the mean FWHM of the night. In the bottom panel of the
Figure we show the relative difference between the FWHM of Qatar-1 and
the FWHM of eight reference stars within the field of view of HSO. We
used two times the standard deviation of the data points as an upper
limit to assess the dispersion of the FWHM. Equivalently, a similar
procedure was repeated for the intensity peaks.

To model the residual modulation in the light curves caused by changes
in the photometric conditions, for each epoch the code generates the
random number $\mathbb{C}$, close to 0.5. Then, for the target and the
$i=1,\ldots, n$ reference stars, the code produces $n+1$ values of $\epsilon_i$
and $n+1$ values of $a_i$. With Eq.~\ref{eq:AB} the intercepts $b_i$ are
determined, and by means of Eq.~\ref{eq:AmpPeak} the peak intensities
for the $n+1$ stars are finally obtained. The resultant modulation is
given by:
\begin{eqnarray}
  FWHM_{\mathrm{mod}} = F_\star/(\Pi_{i=1}^n F_{\mathrm{ref},i})^{1/n}\;,
\end{eqnarray}
The primary transit light curves are then modified by the
estimated $FWHM_{\mathrm{mod}}$.

\subsubsection{Effects associated to poor observing conditions}
\label{sec:h}

A photometric night is neither defined by the brightness of the sky,
nor by its extinction value. Since an increase of the sky brightness
can be compensated by longer exposure times, and a low extinction
coefficient only means that the sky is fairly transparent, what
defines a photometric night is the stability of the sky conditions as
the night evolves. Obviously, not all nights are photometric. Cirrus
cloud formation (i.e. thin clouds holding ice crystals located at
altitudes above 5000 m) are a common phenomenon. They can be easily
noticed during the day or during the night under the presence of the
Moon, when Moonlight is reflected by the ice crystals within the
clouds. However, cirrus can go unnoticed during dark nights.

When sky conditions are far from optimal, true flux levels of stellar
sources cannot be properly measured. They are modulated by the
continuous fluctuations dominating the sky conditions. Generally, sky
variability translate into the data in two forms. In the first
case, the scatter of the data are correlated with the night quality. In
the second case, when the clouds are inhomogeneous throughout the field of
view and change their position rapidly, the light curves show data
points clearly outside the normal data distribution. In addition,
observatories can be light-polluted. This dramatically reduces the
visibility of the stars and enhances, in turn, the effects associated
to fluctuations of the night sky.

Due to the random nature of this effect, we approach its modeling by
analyzing real HSO and OAM Qatar-1 data. To this end, we considered 23
observing nights and counted how many points were observed away from
the normal distribution. A given photometric point was considered an
outlier if it was more than \mbox{$\pm 2 \sigma$} displaced, being
$\sigma$ the natural scatter of the data, estimated from each residual
light curve. From the analyzed HSO and OAM light curves of Qatar-1,
among a total of 2651 data points, 136 were outside the \mbox{$\pm 2
  \sigma$} limit, which corresponds to 5.13\% of the total
datapoints. Therefore, for each synthetic light curve the code
randomly selects between 3\% to 7\% of synthetic data points to be
placed as outliers. To produce the shift, we calculated a local
standard deviation, taking into account only the flux measurements in
the vicinity of the randomly selected points, in order to correlate
the amplitude of the jump with the actual local dispersion of the
data. We then randomly increase or decrease the position of the points
from two up to three times the local standard deviation. Their
corresponding error bars are increased by the same amount. We don't
consider increasing the jump further, because it is of common use to
filter outliers above \mbox{$\pm 3 \sigma$} \citep[see
  e.g.,][]{Moutou2004}.

\subsection{Photometric errors}
\label{sec:i}

\begin{figure}[ht!]
  \centering
  \includegraphics[width=.5\textwidth]{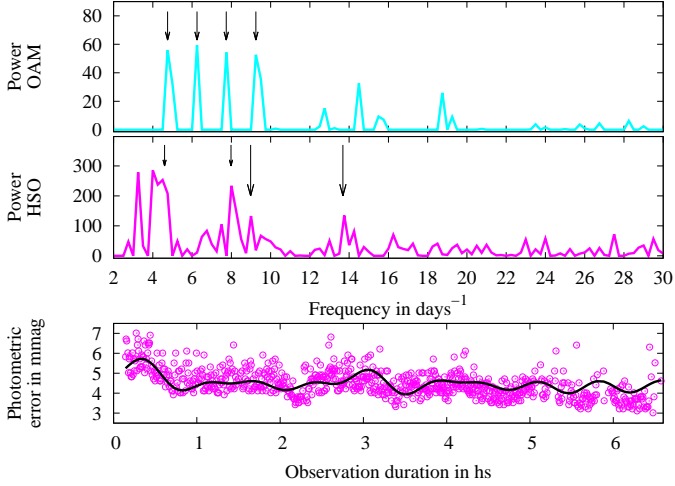}
  \caption{\label{fig:errtreat1} From top to bottom: most
    representative sky fluctuations for OAM (cyan line) and HSO (pink
    line). The bottom plot corresponds to the errors for one observing
    night at HSO. In black, considering the four main frequencies,
    $\mu$ and $\sigma$, we fitted only the phases to the light curve
    (black line) to show the goodness of our approach.}
\end{figure}

Photometric errors are usually provided by a photometric reduction
task. However, reduction tasks do not account for systematic effects
over the photometry. As a consequence, the photometric errors are
slightly underestimated \citep[see e.g.,][for IRAF's
  case]{Gopal-Krishna1995}. For this reason, we did not use the
magnitudes of the errors of real photometric data to create the
synthetic ones, but analyzed them to quantify their dependence with
airmass and the frequency at which they vary. From real error
measurements we found that a linear correspondence with airmass can
properly represent how do error magnitudes change as the stars cross
the sky. Furthermore, due to continuous changes in the sky conditions
the photometric errors also fluctuate. To estimate the frequencies,
$\nu_k$, at which the sky tends to vary more often, we run a
Lomb-Scargle periodogram \citep{Lomb,Scargle,LombScargle} over the
errors computed with IRAF's \textsc{phot} task over OAM and HSO
data. To this end, we analyzed 25 observing nights at HSO spanning two
years, and 10 nights at OAM covering one year. Once individual
periodograms were calculated, we added them up and used the four main
peaks that are more relevant for a transit observation duration (of
the order of a couple of hours) to describe the fluctuations of the
sky (Figure~\ref{fig:errtreat1}).

For our synthetic data as error magnitude, $\epsilon$, we used the
standard deviation of the residual light curve once the stellar
intrinsic and instrumental variability were added to the light curves
(see Sections~\ref{sec:d} to \ref{sec:i}). Using $\epsilon$, the
frequencies $\nu_k$ at which the sky tends to vary more often, and a
phase value $\phi_k$ randomly selected between 0 and 1, the final
photometric errors $\hat{\epsilon_j}$, for each observation $j$, are
estimated as follows:

\begin{equation}
  \hat{\epsilon_j} = \epsilon_j \prod_{k=1}^m\ \sin\ [2\pi(\nu_k\ BJD_j + \phi_k)]\; ,
\end{equation}

\noindent with m = 1,$\cdots$,4.

\subsection{Final light curves}

\begin{figure}[ht!]
  \centering
  \includegraphics[width=.5\textwidth]{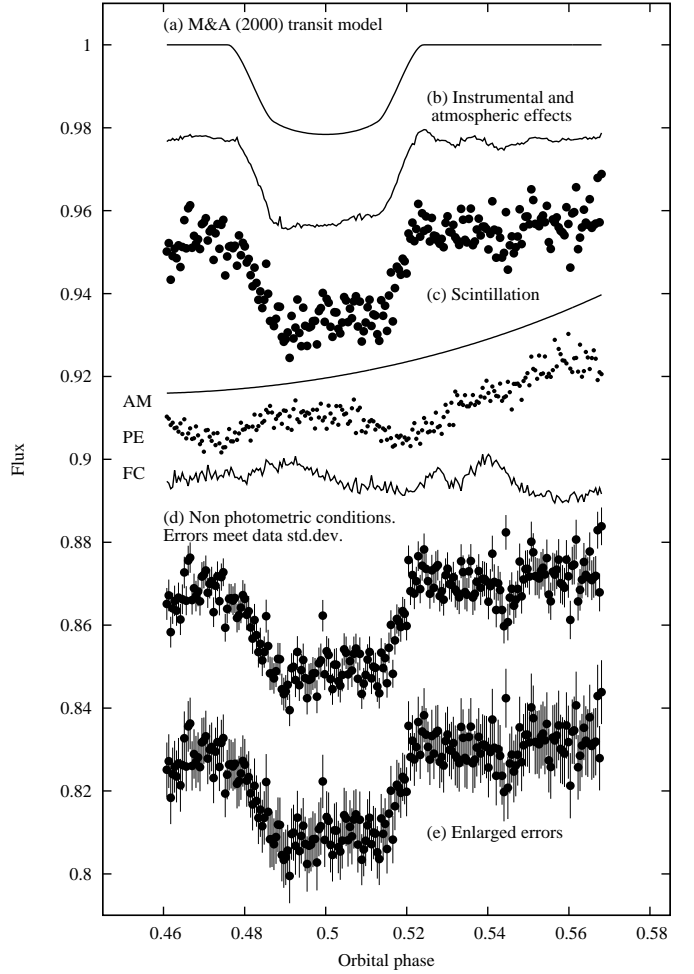}

  \caption{\label{fig:evolution} From top to bottom: (a) Initial
    transit light curve (Sect.~\ref{sec:Agol}). It defines the
    duration of the observation. From now on, all the previous effects
    are considered in the next light curve. (b) Instrumental
    (Sect.~\ref{sec:d}) and atmospheric (Sect.~\ref{sec:e},
    Sect.~\ref{sec:f}) inhomogeneities. (c) Scintillation
    (Sect.~\ref{sec:g}). (d) Effects related to poor observing
    conditions and seeing-related variability (Sect.~\ref{sec:npc},
    Sect.~\ref{sec:npc_seeing}, and Sect.~\ref{sec:h}) plus
    photometric errors (Sect.~\ref{sec:i}). The average value of the
    error bars has been scaled up to meet the standard deviation of
    the light curve. (e) Same as (d), but the error bars have been
    enlarged by $\beta$ (Sect.~\ref{sec:CorrNoise}). The next
    quantities have been scaled and shifted to meet the plot. AM:
    airmass trend during observations. PE: time-variability of the
    photometric errors. FC: integrated counts in the synthetic flat
    field following the computed (x,y) pixel shifts.}
\end{figure}

Figure~\ref{fig:evolution} shows how one particular synthetic light
curve evolves, when the correlated noise sources are sequentially
added to it. From top to bottom we show the \cite{MandelAgol2002}
transit model (a), the latter when the instrumental and environmental
effects are being added to the transit model (b), how scintillation
reflects into the light curve (c), and how non-photometric conditions
impact the data (d). In this case, the error bars have been scaled so
that their averaged value can meet the standard deviation of the
data. The final light curve, with its photometric errors enlarged by
its corresponding $\beta$ value (see Section~\ref{sec:CorrNoise}), can
be found under (e). These are the primary transit light curves from
which the mid-transit times will be retrieved.

\section{Testing our light curves: Real vs. synthetic data}
\label{sec:test}

\begin{figure}[ht!]
  \centering
  \includegraphics[width=.5\textwidth]{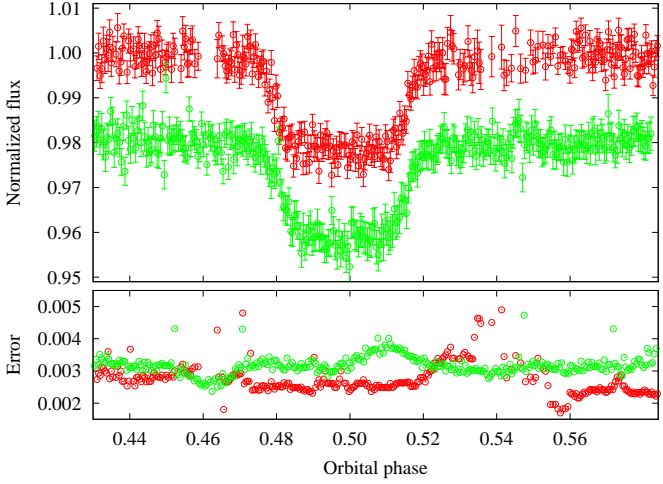}
  \caption{\label{fig:transit_final} Real (red) and synthetic (green)
    light curves of Qatar-1, produced using Oskar L\"uhning telescope
    and Johnson-Cousins R filter. Bottom panel shows how error bars
    change due to non-photometric conditions.}
\end{figure}

Figure~\ref{fig:transit_final} shows real versus synthetic transits of
Qatar-1. In both cases, the observations were performed and simulated
using Johnson-Cousins R filter and Oskar L\"uhning Telescope at HSO.

As initial test, both light curves are visually comparable. The most
important difference (and advantage) in favor of synthetic light
curves is the fact that correlated noise sources are completely
known. For real light curves, however, one can only estimate how much
are they affected by red noise, but not exactly why and how. In order
to properly test the similitude between real and synthetic light
curves, we performed a more concise analysis described in detail under
the following Sections.

\subsection{Comparing time-correlated noise structure}
\label{sec:CorrNoise}

\cite{Pont2006,Carter2009} and references therein investigated how
time-correlated noise affects the estimation of the orbital
parameters. To quantify how dominated are our synthetic light curves
by red noise, we reproduced their analysis as follows: by subtracting
the primary transit model to each light curve, we produced light curve
residuals. We then produced M equally-large bins, varying M between 1
and 40 depending on the available data points per transit light curve
and calculated N, the average value of data points per bin, which
accounts for unevenly spaced data. If residuals are not affected by
red noise, they should follow the expectation of independent random
numbers \citep{Winn2008}:

\begin{equation}
 \sigma_N = \sigma_1 N^{-1/2}[M/(M-1)]^{1/2}\; ,
\end{equation}

\noindent where $\sigma_1$ is the sample variance of the unbinned
data. $\tilde{\sigma_N}$ is the standard deviation of the binned data:

\begin{equation}
 \tilde{\sigma_N} = \sqrt{\frac{1}{M}\sum_{i = 1}^{M}(\mu - \mu_i)^2}\; ,
\end{equation}

\noindent where $\mu_i$ corresponds to the mean value of the residuals
inside each bin, and $\mu$ to the mean value of the means $\mu_i$. If
correlated noise is present, then each $\tilde{\sigma_N}$ will differ
by a factor $\beta_N$ from their expectation $\sigma_N$. By averaging
$\beta_N$ over timescales that are judged to be important for transit
observations (ingress or egress duration), the parameter $\beta$ can
be estimated. $\beta$ accounts for the strength of correlated noise in
the data. For \mbox{Qatar-1}, the time between first and second
contact (or equivalently, the time between third and fourth contact)
is \mbox{$\Delta n \sim$ 15 min}. To estimate $\beta$, we averaged
individual $\beta_N$'s calculated out from bins with sizes 0.8, 0.9,
1.0, 1.1 and 1.2 times $\Delta n$.

\begin{figure}[ht!]
  \centering
  \includegraphics[width=.5\textwidth]{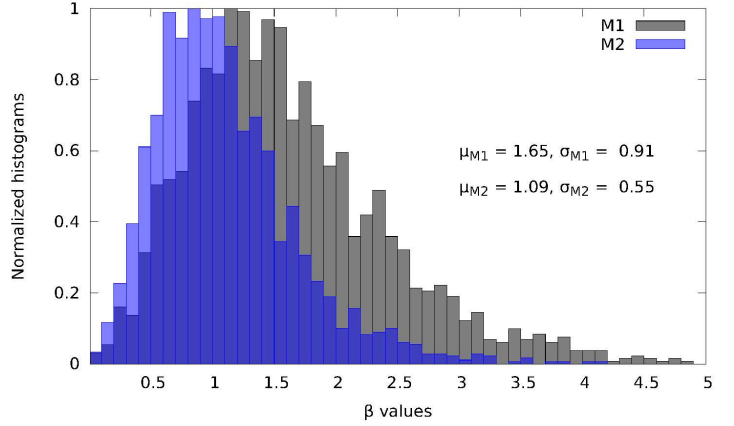}
  \caption{\label{betaStats}Histogram of the strength of correlated
    noise $\beta$ considering raw (blue) and normalized (black)
    synthetic light curves.}
\end{figure}

Figure~\ref{betaStats} shows two normalized histograms of the $\beta$
values that were computed from the available synthetic light curves
produced after 2$\times$35 runs of our code. In more detail, the
$\beta$'s were calculated from the residual synthetic light curves,
which in turn were obtained fitting to the synthetic data a transit
model in simultaneous to a time-dependent low-order polynomial (M1,
black), and a transit model in simultaneous to a linear combination of
some time-dependent environmental and instrumental quantities such as
airmass, seeing, and integrated flat counts (M2, blue). For a more
detailed description about the normalization process, we refer the
reader to Section~\ref{sec:Recovering}. Generally, \mbox{$\beta$ = 1}
corresponds to data sets free of correlated noise. $\beta$ values
smaller than 1 are due to statistical fluctuations and are neglected
in this work. In other words, if a $\beta$ value turns out to be
smaller than 1, the error bars are left unchanged. The most
representative values of the histograms and their scatter,
\mbox{($\mu$, $\sigma$)}, are added to the plot. As the histograms
reveal, M2 data detrending appears to take care more properly of
systematics in the data, since their retrieved $\beta$'s appear to
cluster closer to 1.

\begin{figure}[ht!]
  \centering
  \includegraphics[width=.47\textwidth]{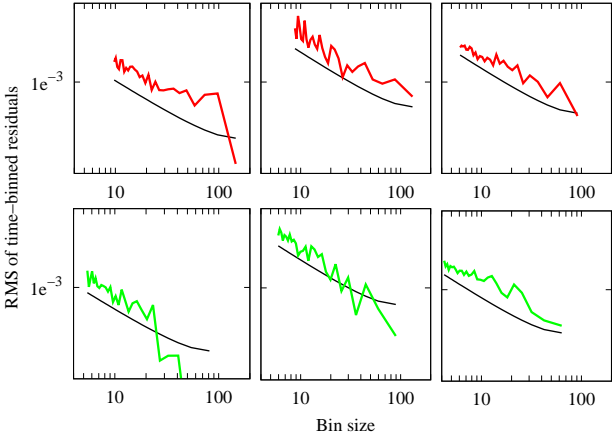}
  \caption{\label{corrNoise} RMS of time-binned residuals as a
    function of bin size for real (red and top) and synthetic,
    M1-normalized (green and bottom) light curves.}
\end{figure}

To test the $\beta$-values obtained from our synthetic data, we
compared them to $\beta$-values obtained from real photometry. For a
quick comparison of the noise structure, Figure~\ref{corrNoise} shows
the results of our correlated noise analysis for the longest three
nights of Qatar-1 real data on top (red lines), and three synthetic
light curves with similar duration and cadence on bottom (green
lines). In all cases, black lines show how residuals should behave in
absence of red noise. Red and green lines represent the variance of
the binned data for HSO real and synthetic light curves, respectively,
as a function of the bin size. As expected, the larger the bin size,
the smaller the RMS. For some of our available Qatar-1 primary transit
light curves we estimated $\beta$ by averaging $\beta_N$ over the same
5 bin sizes already stated. Comparing the synthetic $\beta$-value
distributions against the ones obtained from real data, $\sim$90\% of
our synthetic light curves present the same amount of correlated
noise. Considering that the number of synthetic light curves
significantly exceeds our observations, the correlated noise is indeed
comparable.

\subsection{Comparing autocorrelation signals}

In statistics, autocorrelation occurs when residual error terms from
observations of the same variable at different times are
correlated. If residuals are dominated by Gaussian white noise, then
the normalized autocorrelation of the residuals follows a normal
distribution with mean $\mu$ = 0 and dispersion \mbox{$\sigma$ = 1/N},
being N the number of data points. Ideally, for white noise most of
the residual autocorrelation signal should fall within 95\% confidence
bands around the mean. If the autocorrelation signal of a given data
set doesn't behave as mentioned, then the data accounts for correlated
noise.

\begin{figure}[ht!]
  \centering
  \includegraphics[width=.5\textwidth]{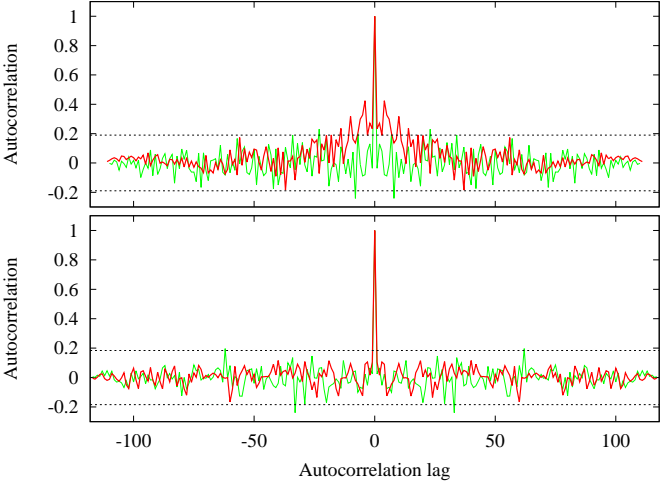}
  \caption{\label{ACrealVSwhite} Autocorrelation function for real
    residual light curves (red) and simulated white noise (green) as
    comparison. The 95\% confidence band is indicated with
    black-dashed horizontal lines. {\it Top:} real data with
    correlated noise. {\it Bottom:} Autocorrelation function falls
    inside the 95\% confidence band.}
\end{figure}

Figure~\ref{ACrealVSwhite} shows an example of the difference in the
residual autocorrelation that exists among real photometry of Qatar-1
obtained during two different nights. As a comparison, the
autocorrelation for simulated residuals affected only by Gaussian
white noise is plotted in green, along with the 95\% confidence band
indicated in black-dashed lines. The autocorrelation function for two
HSO observing nights is plotted in red. On top, the real light curve
is affected by correlated noise, since the central part of the
autocorrelation function clearly escapes the 95\% confidence band. On
bottom, correlated noise appears to be negligible.

\begin{figure}
  \centering
  \includegraphics[width=.5\textwidth]{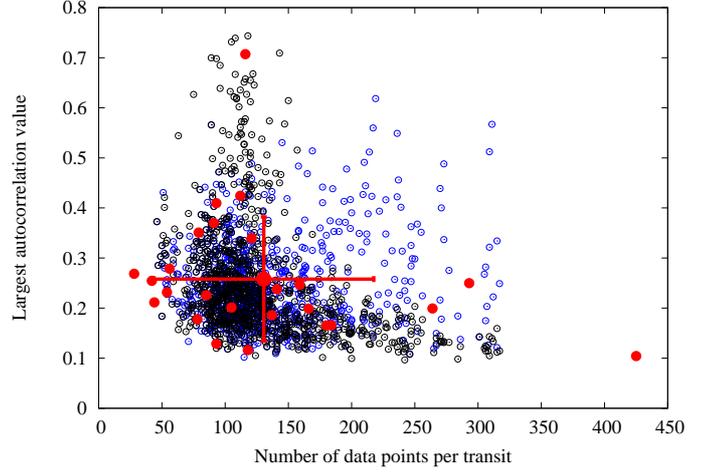}
  \caption{\label{autocorr} Largest autocorrelation value (LAC) for
    synthetic data calculated from residuals obtained subtracting to
    the synthetic light curves the transit model only (blue) and a
    transit model times a second order time-dependent polynomial
    (black). Over-plotted in red, the LACs for 25 real observations
    that were obtained using HSO and Johnson-Cousins R filter.}
\end{figure}

Furthermore, we compared the autocorrelation structure between real
and synthetic data sets (Figure~\ref{autocorr}). The autocorrelation
function was calculated from the residual light curves, obtained after
the primary transit model was subtracted. To compare real to synthetic
light curves we carried out the following analysis: we first
calculated the autocorrelation function of real photometric data, and
plotted the largest autocorrelation value $AC_{max,real}$ as a
function of the data point number (red filled circles). Since it only
takes to validate $AC_{max,real}$ against the 95\% confidence band to
estimate if the light curves are indeed affected by red noise, we
considered sufficient to use $AC_{max,real}$ to compare both sets. As
expected, there is a trend that follows smaller $AC_{max,real}$ values
for larger N's. Finally, we estimated \mbox{$\mu_{AC_{max,real}} \pm
  \sigma_{AC_{max,real}}$} and \mbox{$\mu_N \pm \sigma_N$}.

We then repeated the same process calculating the largest
autocorrelation value from residuals obtained subtracting to the
synthetic light curves the transit model only (blue) and a transit
model times a second order time-dependent polynomial (black). In both
cases, $\sim$80\% of the data points fall within the
1$\sigma_{AC_{max,real}}$, plotted in Fig.~\ref{autocorr} with red
error bars. Taking into account that we count with substantially more
synthetic than real data, the remaining 20\% can be
neglected. Therefore, real and synthetic data seem to present similar
correlated noise structure.

\subsection{Comparison with previous works}

A way to study the impact of systematics over the determination of the
mid-transit times is to produce systematic components with similar
time-scales as the instrumental and environmental systematics, and to
generate synthetic light curves by adding stochastic functions with
similar amplitude and frequency to the real noise present in
photometric data. This approach has been already carried out by other
authors. For example, \cite{Carter2009} created synthetic light curves
which correlated noise was represented by the sum of two uncorrelated
and correlated Gaussian processes, and focused on the impact of this
noise structure over the determination of the mid-transit
times. \cite{Gibson2013} created synthetic data adding up a ``function
noise'' that was build by summing up 100 exponential, Gaussian, and
sinusoidal functions with random parameters, with the main goal to
test the accuracy of the retrieved orbital parameters. Although the
analysis we have in common produce results that do not differ (see
Section~\ref{sec:Recovering}) the main differences between our method
and previous work is that, in our case, the time dependency is
represented more realistically. For example, our method accounts for
noise that improves or degrades as the stars cross the sky (airmass
and seeing dependency), which is observed in real photometric data. We
can also investigate the impact of current detrending techniques into
the determination of the orbital and physical parameters of the system
(see Section~\ref{sec:results}), because we have environmental and
instrumental quantities (e.g., airmass, seeing, and changes of the
centroid position of the star over the CCD) that can be used to
understand the impact of these systematics over transit light
curves. Indeed, in this work we study how much does the precision of
the orbital and physical parameters improve when a certain
normalization is being considered, which would have been impossible to
carry out when only stochastic functions are used to represent the
noise.

\section{Recovering the TTV signal: General aspects}
\label{sec:Recovering}

Once the synthetic light curves are generated (usually between 50 to
70 each run), recovering the TTV signal is the next step to
follow. Before transit fitting begins, we visually inspect the
generated light curves, removing those presenting extremely poor
transit coverage, scatter larger than the transit depth, and
highly affected by correlated noise. The number of deleted light
curves clusters around 5-10 per run. Then, the TTV recovery goes as
follows:

\begin{itemize}
\item To obtain good estimates for the system parameters, we select
  the five best light curves according to the following criteria: they
  should have a good amount of OOT data before and after transit
  begins and ends, respectively, small scatter compared to the transit
  depth, and good cadence. The transits should also be well spread
  along the 400 epochs to retrieve an accurate orbital period, and
  should be divided among the 5 filters and the 3 observatories that
  the code considers (Sect.~\ref{sec:GBO}). It is worth to mention
  that an incorrect selection of transit light curves (i.e., by
  considering primary transits with large scatter, incomplete, or
  strongly affected by correlated noise) leads to {\it very}
  inaccurate orbital parameters. This selection was done by
    visually inspecting the light curves generated over more than 60
    full runs of the code.
\item From the latter sub-sample, we choose the best light curve with
  respect to data scatter and sampling rate. It will be considered as
  the 0$^{th}$ epoch.
\item Together with the transit model
  \citep{MandelAgol2002}\footnote{\url{http://www.astro.washington.edu/users/agol}}
  we simultaneously fit to the data two models accounting for the
  non-transit variability, but separately. In other words, for each
  run we will fit the data twice. To reproduce as best as possible
  current data detrending techniques, in this work we consider a
  low-order time-dependent polynomial (first, second or third order,
  from now on called M1 normalization). The selection of the order is
  carried out light curve by light curve by minimizing the Bayesian
  Information Criterion, \mbox{BIC = $\chi^2$ + k ln(N)}. For the BIC,
  k is the number of fitting parameters, N is the number of data
  points per light curve, and $\chi^2$ is computed from the residuals,
  which in turn are obtained by subtracting to the synthetic data the
  best-fit model with its corresponding time-dependent polynomial. As
  detrending function we also consider a linear combination of airmass
  (AM), seeing (SN), x and y pixel position ($x_{pix}, y_{pix}$) and
  integrated flat counts in those (x,y) values (FC), from now on
  called M2 normalization. All these quantities are provided by the
  code. Thus, in the first case the detrending fitting parameters are
  up to four, while in the second case they are six: the previously
  mentioned ones plus an offset. In short, the time dependency of the
  normalization functions and the fitting parameters look as follows:

  \begin{equation}
    M1(t) = a_3 t^3 + a_2 t^2 + a_1 t + a_0\;,
  \end{equation}

  \begin{equation}
    M2(t) = c_0 + c_1 \mathrm{AM} + c_2 \mathrm{SN} + c_3 x_{\mathrm{pix}} + c_4 y_{\mathrm{pix}} + c_5 \mathrm{FC}\;.
  \end{equation}

\noindent The simultaneous fitting of the transit model and the
detrending function is carried out in the same fashion not only to
compute the orbital parameters of the system but to obtain the
mid-transit times of the individual light curves.

\item We then proceed to fit the five selected transit light curves by
  sampling from the posterior probability distribution using a
  Markov-chain Monte Carlo (MCMC) approach. From the transit light
  curve we can directly infer the following parameters: the orbital
  period, $P$, the mid-transit time, $T_{o}$, the planet to star
  radius ratio, $p=R_p/R_s$, the semi-major axis in stellar radii,
  $a/R_s$, the orbital inclination, $i$, and the limb-darkening
  coefficients. For our fits we assume a quadratic limb-darkening
  prescription with fixed $u_1$ and $u_2$
  (Table~\ref{tab:limbdarkening}). From now on, these are called
  global parameters.
\item After $5\times10^5$ iterations we discard a suitable burn-in
  ($10^5$ samples) and compute the best fit parameters from their
  posterior distributions (mean and standard deviation as best-fit
  values and errors). The errors for the global parameters are derived
  from the 68\,\% highest probability density or credibility intervals
  (1 $\sigma$).
\item Afterwards, we fit each light curve individually in an
  equivalent fashion as in the two previous steps. To consider the
  existing information in the determination of the individual
  mid-transit times, rather than fixing the orbital parameters to
  their best-fit values we specify Gaussian priors on $a/R_s$, $i$,
  and $p$. Since now the transit light curves are analyzed separately
  the orbital period, $P$, is left fixed to the global best-fit
  value. As previously mentioned, the model fitted to the data is the
  product between the transit model and M1 or M2. Before the
  individual light curves are fitted, we calculate the $\beta$ value
  as specified in Section~\ref{sec:CorrNoise} and we enlarge the
  individual photometric error bars by it. Finally, we obtain the
  best-fit mid-transit times, $T_{o,k}$, along with their error
  estimates that are drown from MCMC chains at 1$\sigma$ levels.
\item To produce the synthetic O--C diagram, we consider the
  ``Calculated'' mid-transits as an integer multiple (epoch number) of
  the global orbital period. ``Observed'' mid-transits are the ones
  individually fitted in the previous step.
\end{itemize}

\begin{figure}[ht!]
  \centering
  \includegraphics[width=.5\textwidth]{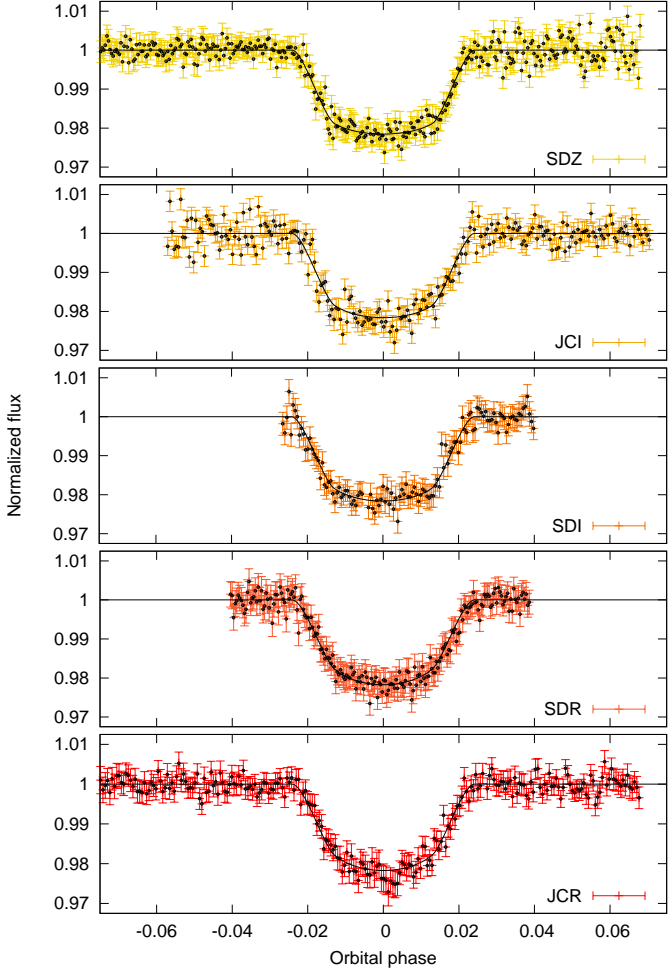}
  \caption{\label{fig:lightcurves3} Sample of synthetic light
    curves. In continuous line, \cite{MandelAgol2002} primary transit
    model for each limb darkening coefficient set.}
\end{figure}

\begin{figure}[ht!]
  \centering
  \includegraphics[width=.5\textwidth]{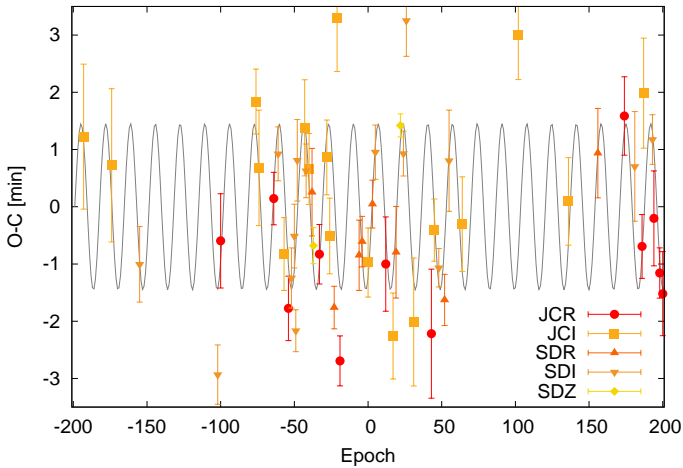}
  \caption{\label{fig:OC_example} Synthetic O--C diagram. The
    mid-transit shifts are color-coded according to the randomly
    selected filter. The timing variation added by the code to each
    mid-transit time is plotted with gray continuous line. Clearly,
    some points scatter outside the true TTV signal.}
\end{figure}

Figure~\ref{fig:lightcurves3} shows five synthetic light curves
previously normalized by a time-dependent polynomial, one for each
available filter. The data quality and their duration vary
considerably. Light curves of this kind, combined all together, will
be the ones used to perform the TTV analysis. In addition,
Figure~\ref{fig:OC_example} shows one of the many synthetic O--C
diagrams, obtained from the previously described procedure. The
observed ``grouping'' of data points in the O--C diagram is caused by
visibility effects, another feature observed in real transit
follow-ups which can have an impact in the determination of the
perturbers orbit if not treated properly (see
Section~\ref{sec:periodogram}).

\section{Recovering the TTV signal: results}
\label{sec:results}

\subsection{Determination of the global parameters for different normalization strategies}
\label{sec:global}

It is not news that the determination of the individual mid-transit
times strongly depend on the normalization of the photometric data
\citep{Winn2008,Gibson2009}, specially when incomplete light curves
are taken into account. The choice of normalization has also an impact
over the determination of the orbital and physical parameters, which
in turn can affect the value of the mid-transit time. Therefore, we
investigated the impact of the M1 and M2 normalizations by
investigating the accuracy of the semi-major axis, the inclination,
the orbital period, and the planet-to-star radius ratio.

\begin{figure}[ht!]
  \centering
  \includegraphics[width=.5\textwidth]{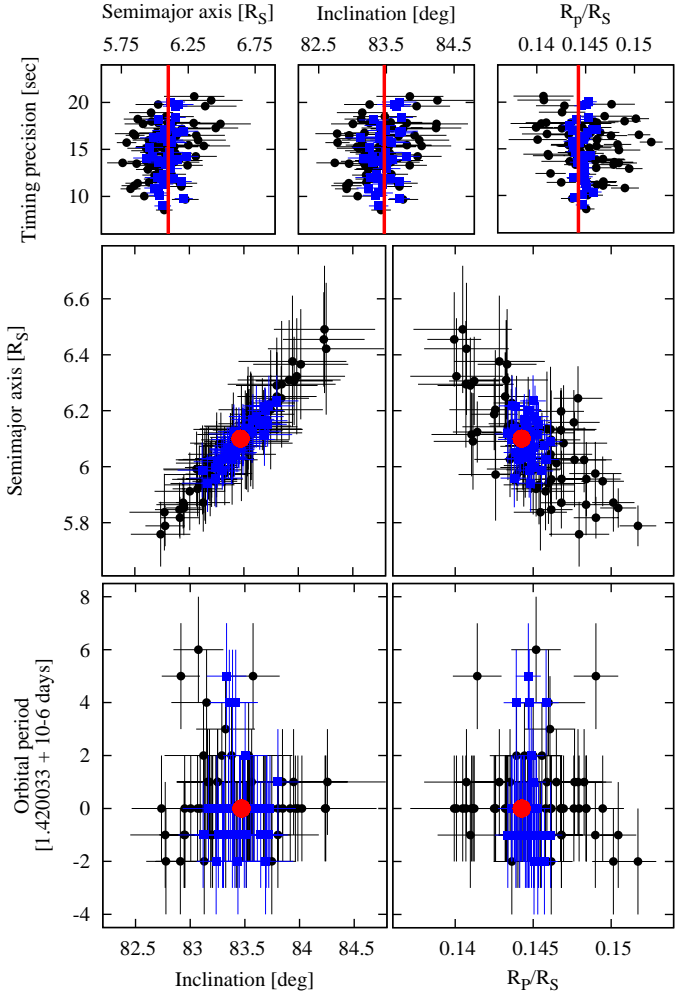}
  \caption{\label{fig:GlobalParams} Global parameters obtained
    normalizing the synthetic data via M1 (black circles) and M2 (blue
    squares). The red points indicate the values used by the code to
    produce the synthetic data, equivalently to the red vertical
    lines. The top three panels show the timing precision versus the
    derived transit parameters.}
\end{figure}

The best-fit global parameters obtained as described in
Section~\ref{sec:Recovering}, along with their errors, are plotted in
Figure~\ref{fig:GlobalParams}. Black circles and 1$\sigma$ errors
correspond to the global parameters obtained by means of synthetic
data fitted against a transit model and M1. Blue squares and their
respective 1$\sigma$ errors were obtained detrending the data using
M2. The initial orbital parameters used by the code to produce the
synthetic data are indicated in the plot by a red dot. The four plots
in the lower part of the Figure reveal two obvious features:
parameters obtained by M2 normalization have a significant less
scatter than the ones obtained from M1. In consequence, they show more
consistency with the values used as input, and full consistency when
errors are at 2$\sigma$ level. The sub-plot from the upper-left part
of Fig.~\ref{fig:GlobalParams} shows the already known correlation
between the semi-major axis and the orbital inclination via the impact
parameter $b = a\cos(i)$. Comparing both sets of solutions, these
values reveal that not only the normalization has indeed an impact on
the determination of the orbital and physical parameters, but the
choice of normalization as well. Thus, the M2 normalization appears to
better account for systematics and, in consequence, produces more
accurate and consistent transit parameters.

The three smaller panels on top of the Figure correspond to the timing
precision (errors on the 0$^{th}$ epoch at 1$\sigma$ level) versus the
derived semi-major axis, $a$, the inclination, $i$, and the
planet-to-star radius ratio, $R_\mathrm{P}/R_\mathrm{S}$. Although the
mid-transit time (and in consequence its precision) should not depend
on the physical transit parameters, it is not (always) the case, as
observed in the Figure. When the M2 is used as detrending function
there is no strong correlation between the timing precision and the
precision of the transit parameters. However, when the data is
detrended by a low-order time-dependent polynomial (M1) there is a
strong correlation between the physical parameters and the timing
precision. As the Figure shows, the uncertainties of the parameters
increase as the timing precision decreases, as well as their scatter
around the input value, creating in some cases inconsistency. It looks
like the normalization procedure affects the transit parameters.

\subsection{Significance of timing offsets: a more conservative approach}
\label{sec:offsets}

\begin{figure}[ht!]
  \centering
  \includegraphics[width=.5\textwidth]{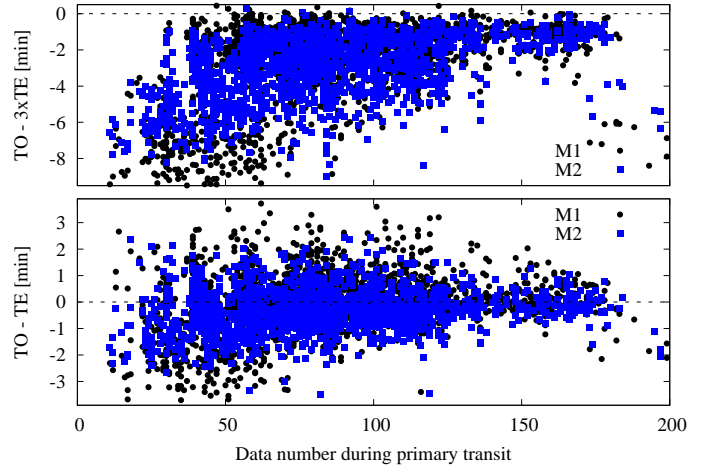}
  \caption{\label{fig:transitDur} {\it Bottom:} absolute value of the
    timing offset (TO) minus the timing error (TE) in minutes. {\it
      Top:} TO minus three times TE. Black circles correspond to M1
    normalization and blue squares to M2. The plot has been produced
    from the analysis of 35 synthetic O--C diagrams.}
\end{figure}

To estimate how much are the mid-transit times affected by the transit
duration, we run the code 35 more times but shutting the TTVs
off. Transits produced in this fashion were analyzed as described in
previous Sections. Therefore, if any significant timing variability is
present, this should be caused by systematics not properly taken care
of. From these transits we computed the timing offset (TO) which is
the absolute value of the difference between the observed timing shift
and their expected shift (in this case 0), and subtracted to it their
corresponding timing error, TE. A negative \mbox{TO - TE} would
correspond to a TTV consistent with zero. Equivalently, a positive
\mbox{TO - TE} would indicate a significant timing offset. The bottom
panel of Figure~\ref{fig:transitDur} shows our results, when M1 and M2
normalizations are implemented (black circles and blue squares,
respectively). \mbox{TO - TE}'s are plotted against the number of data
points during primary transit, but the same exercise was performed for
the number of OOT data points, the standard deviation of the residual
light curves, and the transit coverage. The top panel of the Figure
shows how an increment of the error bars to a 3$\sigma$ level gives
timing offsets completely consistent with zero, an exercise that we
would suggest the reader to carry out when assessing TTVs from
ground-based observations. To quantify the correlation of the
mid-transit times that would be consistent with TTVs (i.e., positive
\mbox{TO - TE}) we used the Pearson correlation coefficient $r$:

\begin{equation}
  r_{xy} = \frac{\sum_{i=1}^n\ (x_i - \mu_x) (y_i - \mu_i)}{[\sum_{i=1}^n\ (x_i - \mu_x)^2 \sum_{i=1}^n\ (y_i - \mu_y)^2]^{1/2}}\; ,
\end{equation}

\noindent for $x$ number of data points during primary transit and $y$
the timing residuals. \mbox{$r_{M1} = -0.25$} and \mbox{$r_{M2} =
  -0.19$} confirm the existence of the correlation, which is
observable even by visual inspection. Similar results were obtained
analyzing the timing residuals against OOT data points and transit
coverage. Thus, as previously observed by other authors we caution the
use of incomplete light curves or poorly sampled primary transits to
carry out TTV studies. This also reveals that TTVs derived from
ground-based observations with amplitudes of the order of some minutes
would be the consequence of either an improper treatment of the
systematics, or an underestimation of the timing errors
\citep{Raetz2015} rather than caused by the interaction between two
planets. Therefore, in this case we caution to make any statements
about the detectability of TTVs.

\subsection{Scatter in the O--C diagram: How much do systematics contribute?}

When a periodogram analysis of the O--C diagram reveals a peak, it is
of common practice to use the scatter in the O--C diagram to
characterize the mass and orbital separation of a hypothetical
perturbing planet \citep[see
  e.g.,][]{Adams2010b,Adams2011,vonEssen2013,Awiphan2016}. To this
end, different dynamical scenarios are considered and analyzed
\citep[for example, an inner perturber, an outer perturber, or two
  bodies in mean-motion resonances,][]{Agol2005}. In each dynamical
configuration, the semi-major axis and the orbital eccentricity vary
within a range of possible values. At each step, the scatter of the
theoretical O--C diagram is computed, and compared to the observed
one. This procedure is also repeated considering different masses for
the perturbing body.

As seen in Section~\ref{sec:offsets}, poor primary transit coverage
can yield to a considerably large timing offset that is completely
independent of any TTV. Therefore, to understand whether unaccounted
correlated noise sources lead to under- or over-estimations of the
characteristics of the perturbing body, we analyzed again the scatter
of the O--C diagrams for TTVs set on. In detail, we compared the
scatter of the synthetic O--C diagram to the first order mean-motion
resonance scenario. For each one of the normalization strategies we
computed the observed standard deviation of each synthetic O--C
diagram:

\begin{equation}
  \sigma_{OC,synth} = \left[ \frac{1}{N_{OC}-1}\sum_{k=1}^{N_{OC}}[T_{o,k} - (T_o + P n_{orb,k})]^2 \right]^{1/2}
\end{equation}

\noindent where $N_{OC}$ is the number of light curves that the code
generated in a given run, minus the ones that were deleted after
visual inspection (Section~\ref{sec:Recovering}). $n_{orb,k}$
corresponds to the orbit number with respect to the zeroth epoch, and
$T_o$ and $P$ are the best-fit mid-transit time for the zeroth epoch
and the orbital period, respectively. If two planets coexist in mean
motion resonance, as pointed in Section~\ref{sec:Agol} the
perturbation term $PT(k)$ added to the unperturbed mid-transit times,
for a given $k$ epoch, has the following expression:

\begin{equation}
  PT(k) = \delta t_{max} \sin[2\pi P_{Trans}(k-1)/P_{lib}]\; .
\label{eq:deltat}
\end{equation}

\noindent From Eq.~\ref{eq:deltat} we can compute the theoretically
expected scatter:

\begin{eqnarray}
  \sigma_{model} &=& <(PT(k))^2>^{1/2} \nonumber \\
               &=& <(\delta t_{max} \sin[2\pi P_{Trans}(k-1)/P_{lib}])^2>^{1/2} \nonumber \\
               &=& (\frac{<\delta t_{max}^2>}{2})^{1/2} \nonumber \\
               &=& \frac{\delta t_{max}}{\sqrt{2}}\; .
\label{eq:SIGMA_MODEL}
\end{eqnarray}

\noindent Therefore, to estimate the perturbers mass by comparing the
observed scatter, $\sigma_{OC,synth}$, to the theoretical one,
$\sigma_{model}$, we require information about the orbital period and
the mass of the transiting planet (which is normally known from
transit photometry and radial velocity measurements), in addition to
the order of the resonance. The only parameter that will vary, while
comparing $\sigma_{model}$ to $\sigma_{OC,synth}$, is the mass of the
perturbing body.

To calculate $\sigma_{model}$ we have to consider that each run
provided slightly different orbital parameters
(Section~\ref{sec:global}). Therefore, to be able to compare the
results obtained at each run we considered the orbital period and the
planetary mass used by the code as input parameters, and relatively
large errors for the mentioned parameters (0.1\% and 10\%,
respectively). \cite{Kipping2010} studied the effects of finite
integration times over the determination of the orbital parameters. If
we consider the error on the mid-transit times that large exposure
times produce, we will be able to define a lower limit on the
amplitude of the TTV that we can realistically detect. Since we are
considering only the case of first order mean-motion resonances
produced by an Earth-sized planet, $j\geq3$ would yield TTV amplitudes
too small to be detected for exposure times of the order of one
minute. Furthermore, j = 1 would produce TTV amplitudes easily to
detect, even by means of these light curves
\citep{Agol2005}. Therefore, it would not be inappropriate to restrict
the resonance order to \mbox{$j = 2$}, if our aim is to be consistent
with the data that we have at hand.

\begin{figure}[ht!]
  \centering
  \includegraphics[width=.5\textwidth]{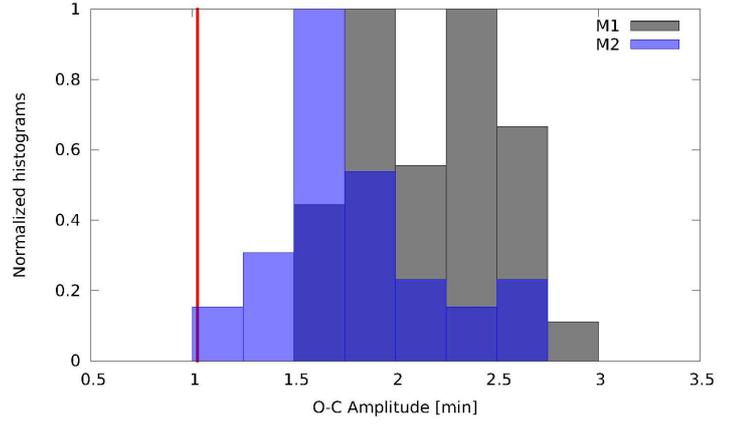}
  \caption{\label{fig:OC_hist} Normalized histograms of the
      amplitude of the synthetic O--C diagrams when the M1 (black) and
      M2 (blue) normalizations are performed. The vertical red line
      shows the expected TTV amplitude produced by the perturber.}
\end{figure}

\begin{figure}[ht!]
  \centering
  \includegraphics[width=.5\textwidth]{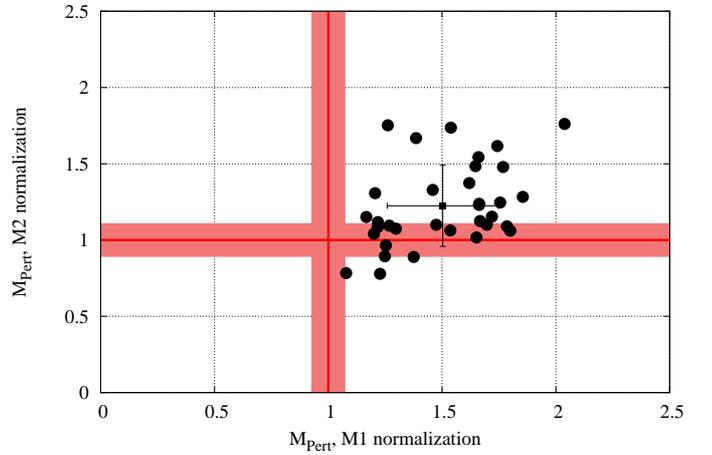}
  \caption{\label{fig:OC_mass} Retrieved mass of the perturber from
    the amplitude of the O--C diagrams in black circles for the M1
    (x-axis) and M2 (y-axis) normalization. The black square with
    error bars shows the mean value and standard deviation of the
    points. In red horizontal and vertical lines, the expected mass of
    the perturber. The red area accounts for errors in the mass and
    orbital period of the transiting planet.}
\end{figure}

Figure~\ref{fig:OC_hist} and Figure~\ref{fig:OC_mass} summarize our
results. The first Figure shows the derived amplitude of 35 synthetic
O--C diagrams that were obtained implementing the M1 (black) and M2
(blue) normalization. While the black distribution has a mean value
around 2.1 minutes, the blue one clusters around 1.5 minutes. The
vertical red line indicates the amplitude in minutes that the input
perturber causes. The second Figure shows the derived perturber masses
assuming a \mbox{j = 2} resonance. Propagating the errors of $P$ and
$m_{Trans}$ previously mentioned allowed us to produce an error
estimate for the perturbers mass. The latter is plotted horizontally
and vertically in red.

After comparing the predicted to the observed scatter, we found two
main results: first, the scatter of the synthetic O--C diagram,
associated to M1 and M2 normalization, seem to overestimate the action
of the perturber. Although the M2 normalization appears to represent
it more adequately, it is only consistent with it in few cases. We
understand this as an improper treatment of the systematics. Second
and most importantly (and consequent to the first case), the planetary
masses obtained from the scatter of the synthetic O--C diagrams are
over-estimated in most of the cases by around 50\% in the case of M1
normalization and around 20\% for the M2 normalization. Therefore,
determining characteristics from the perturber using poor transit
light curves will only provide miss-leading results about the system
under study.

\subsection{Periodogram analysis}
\label{sec:periodogram}

In both continuous and discrete cases, Fourier theory explains how any
function can be represented or approximated by sums of simple
trigonometric {\it and periodic} functions. Given any time series, it
is possible to find sines and cosines with different periodicity,
phases and amplitudes that, when added together, can reproduce the
time series back again.

Regarding TTV studies, once an O--C diagram is produced, the first
natural step is to look for any kind of periodicity associated to the
effects that a perturbing planet might cause on the timings of a
transiting exoplanet. However, correlated noise sources affecting
mid-transit times can disguise true signals. To test how much do
systematic effects not properly accounted for affect the
characterization of TTVs, we analyzed 35 O--C diagrams that were
obtained from synthetic light curves {\it not} affected by transit
timing variations, but affected by every systematic source instead. To
this end we run a periodogram over each O--C diagram, searching for
any leading frequency that could fake a planetary perturber signal but
analyzing the M1/M2 scenarios separately. Once the frequency
corresponding to the maximum power in the periodogram was found, we
computed the spectral window of the O--C diagram to check that this
peak was not caused by the sampling rate itself.  To each one of the
leading peaks we fitted a Gaussian function which mean and standard
deviation correspond to the TTV libration frequency and its
error. Values ranged between 10 and 200 days. For each peak we also
computed the false alarm probability (FAP) \citep{Horne1986} and also
power values corresponding to FAP's of 20\%, 10\% and 1\%. Then, we
counted how many times did these FAP's exceeded the maximum
power. From the 35 O--C diagrams, on average for the M1 and M2
scenarios, we found that the detected maximum peak was higher than
FAP's of 20\%, 10\% and 1\% 14, 7 and 3 times, respectively. It is
worth to mention that all these values should have been
zero. Therefore, we caution the reader to give large FAP limits such
as the ones provided here, when assessing significance levels on TTV
periodicity. In our case, we had to decrease the FAP to 0.1\% to have
all the leading peaks below this level.

\subsection{Quality factor}

\begin{figure}[ht!]
  \centering
  \includegraphics[width=.5\textwidth]{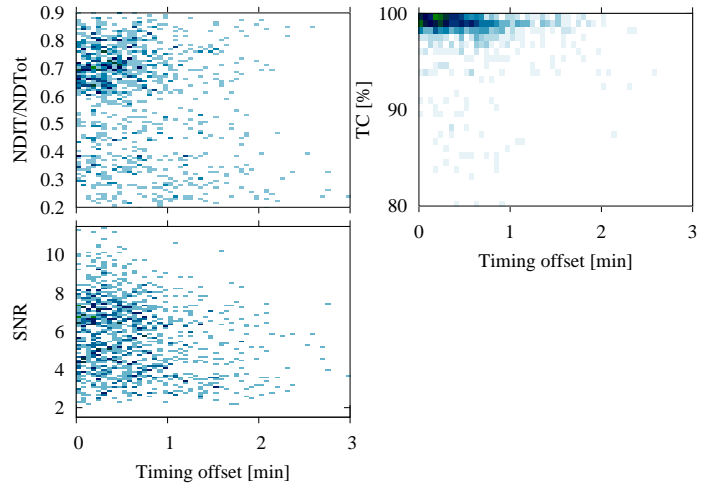}
  \caption{\label{fig:QF} Density maps of light curves that produced
    consistent timings. From left to right and top to bottom the
    number of data points in transit (NDIT) divided by the total
    number of data points (NDTot), the transit coverage, TC, in
    percentage, and the signal-to-noise ratio of the light curve.}
\end{figure}

Rather than suggesting the reader to disregard TTVs obtained from poor
light curves only, we attempt to characterize the quality of the light
curves that, given 1$\sigma$ errors, were consistent with their
expected timing value. In other words, these were light curves that,
although they show a 1-2 minute offset from their expected value, they
were still consistent within errors. This can be seen in
Figure~\ref{fig:QF}. The lower panel of the Figure shows a density map
of the signal-to-noise ratio of the light curves, this is, the transit
depth divided by the standard deviation of the residual light
curves. The top left panel of the Figure shows the number of data
points in transit (NDIT) divided by the total number of data points
per transit light curve (NDTot). The right top panel of the Figure
shows the transit coverage, TC, in percentage. 100\% corresponds to
light curves which primary transit was fully observed. The bluer the
pixel in the density maps, the more light curves were showing these
particular features. Thus, for example light curves with a \mbox{SNR =
  7}, \mbox{NDIT/NDTot = 0.7}, and \mbox{TC = 100\%} would provide
reliable TTVs when the M2 normalization is performed. The three
quantities characterized here can be easily obtained from transit
photometry. Therefore, we suggest the reader to add this as quality
check.

\section{Discussion and conclusion}

In this work we analyze whether current techniques used to detrend
transit light curves taken from ground-based telescopes are suitable
to properly characterize multiplicity in particular transiting systems
via the transit timing variation technique. To this end, we simulated
primary transit observations caused by a hot Jupiter which orbital and
physical configuration mimics a real system, Qatar-1. To these light
curves we artificially added a perturbation in their mid-transit times
caused by an Earth-sized planet in a 3:2 mean motion resonance. The
synthetic data accounts with what we believe are the most significant
sources of light curve deformations: environmental variability
(airmass, atmospheric extinction, and chaotic variability in the sky
conditions during observations) and instrumental variability. We then
tested the quality of our light curves, comparing their noise
characteristics to the ones present in real data.

As already shown by other authors, our results disfavor the use of
incomplete light curves to carry out TTV studies. Furthermore, our
studies show that it is more likely that systematics not properly
accounted for are causing the observed scatter in the O--C diagram
rather than a gravitationally bound exoplanet. This, in consequence,
produces mass estimates of the perturbing body that are a factor of up
to two larger as expected. We also find that transits normalized by a
time-dependent low-order polynomial provide more inaccurate and
sometimes even inconsistent orbital and physical parameters than the
ones derived from a more instrumentally and environmentally-related
detrending function, which includes time-dependent variability such as
changes of airmass, seeing, centroid position and integrated flat
counts. Nonetheless, our results suggest that either both approaches
are insufficient to account for systematics, or error estimates on the
mid-transit times are being underestimated by current statistical
techniques by a factor of up to three. A final inspection of the O--C
diagrams and the light curves associated to each O--C point make us
conclude that when only light curves with close-to-full transit
coverage, good cadence, and large photometric quality are considered
to carry out TTV studies, the derived O--C diagrams appear to be more
consistent with their expected variability. In a future work we will
investigate if the use of Gaussian Process regression can improve the
determination of the perturbers characteristics, which would allow us
to use low-quality transit photometry for TTV studies.


\section{Acknowledgments}

C. von Essen acknowledges funding by the DFG in the framework of RTG
1351, and E. Suarez and A. Ofir for fruitful discussions. Funding for
the Stellar Astrophysics Centre is provided by The Danish National
Research Foundation (Grant DNRF106). The authors acknowledge the
referee for his/her positive feedback and suggestions to improve this
manuscript.

\bibliographystyle{aa}
\bibliography{TTVsim}

\end{document}